\documentclass[a4paper,amsmath,amssymb,pra,floatfix,footinbib,twocolumn,preprintnumbers]{revtex4-1}

\usepackage[dvips]{graphicx}
\usepackage{dcolumn}
\usepackage{bm}
\usepackage{dsfont}
\usepackage{color}
\usepackage[normalem]{ulem} 
\usepackage{url}
\usepackage[colorlinks=true ,citecolor=blue]{hyperref}
\usepackage{amsmath,amssymb}

\def\b{\mathbf}
\def\rm{\mathrm}

\newcommand{\nn}{\nonumber}
\newcommand{\be}{\begin{equation}}
\newcommand{\ee}{\end{equation}}
\newcommand{\bea}{\begin{eqnarray}}
\newcommand{\eea}{\end{eqnarray}}
\newcommand{\mv}[1]{\langle #1\rangle}
\newcommand{\f}{\frac}

\newcommand{\ket}{\rangle}
\newcommand{\ra}{\rightarrow}

\begin{document}

\title{Two-body physics in the Su-Schrieffer-Heeger model}

\author{M. Di Liberto$^{1}$, A.~Recati$^{1,2}$, I.~Carusotto$^{1}$, and C.~Menotti$^{1}$}
\affiliation{
$^1$INO-CNR BEC Center and Dipartimento di Fisica, Universit\`a di Trento, 38123 Povo, Italy\\
  $^{2}$ Arnold Sommerfeld Center for Theoretical Physics, Ludwig-Maximilians-Universit\"at M\"unchen, 80333 M\"unchen,
  Germany}

\date{\today}

\begin{abstract}
We consider two interacting bosons in a dimerized Su-Schrieffer-Heeger
(SSH) lattice.  We identify a rich variety of two-body states. In particular, 
for open boundary conditions and moderate interactions, edge bound
states (EBS) are present even for the dimerization that does not
sustain single-particle edge states. Moreover, for large
values of the interactions, we find a breaking of the standard
bulk-boundary correspondence. Based on the mapping of two interacting particles in one
dimension onto a single particle in two dimensions, we propose an experimentally
realistic coupled optical fibers setup as quantum simulator of the two-body SSH
model. This setup is able
to highlight the localization properties of the states as well as
the presence of a resonant scattering mechanism provided by a bound state that 
crosses the scattering continuum, revealing the closed-channel population in real time
and real space.
\end{abstract} 


\maketitle

\section{Introduction}

In a perfectly periodic system, states outside
the allowed bands can appear for both attractive and repulsive
interactions when composite objects are formed
\cite{Hubbard1963,MattisRMP}.  The existence of ``exotic" repulsive
bound pairs, a.k.a. doublons, has been directly observed for the first
time ten years ago by implementing a single-band Hubbard model with
ultra-cold Bose gases in an optical lattice \cite{winkler2006}.  The
study of doublons has been extended to, e.g., long range interacting
particles \cite{Valiente2009,Valiente2010,Longhi2012}, two-channel
models \cite{Nygaard2008,Valiente2011}, superlattices
\cite{Valiente2010b} and spinor gases \cite{Menotti2016}.
Aside from presenting behaviours and stability properties interesting
by themselves \cite{Esslinger}, doublons deeply affect the dynamics of
the system.  For instance very recently it has been shown that the
presence of doublons favours many-body localization in disordered
\cite{BlochMBL2015} or extended \cite{Barbiero2015} Hubbard models.

On the other hand, any real crystal is made of a bulk and a
surface. The study of how surfaces modify the spectrum of a particle
in a finite crystal started with the seminal papers by Tamm
\cite{Tamm32} and Shockley \cite{Shockley39}. They pointed out the
existence of localized states at the surface with energy outside the
allowed energy bands.  Such surface states can play an important role
in the transport properties. Particular attention has been devoted
in the recent years to their characterization in the so-called
topological insulator materials \cite{KaneRMP10}. The {\it
  bulk--boundary correspondence} provides a link between the presence
and number of in-gap edge states and the topological invariants of the
bulk crystal. The most famous example is the chiral state on the edge
of a two-dimensional integer quantum Hall system (see
e.g. \cite{Tong}). While most of the above mentioned surface states
are well explained by single-particle band theory, the physics becomes
much more intriguing in the presence of strong inter-particle
interactions. 

\begin{figure}[!tbp]
\includegraphics[width=1.\columnwidth]{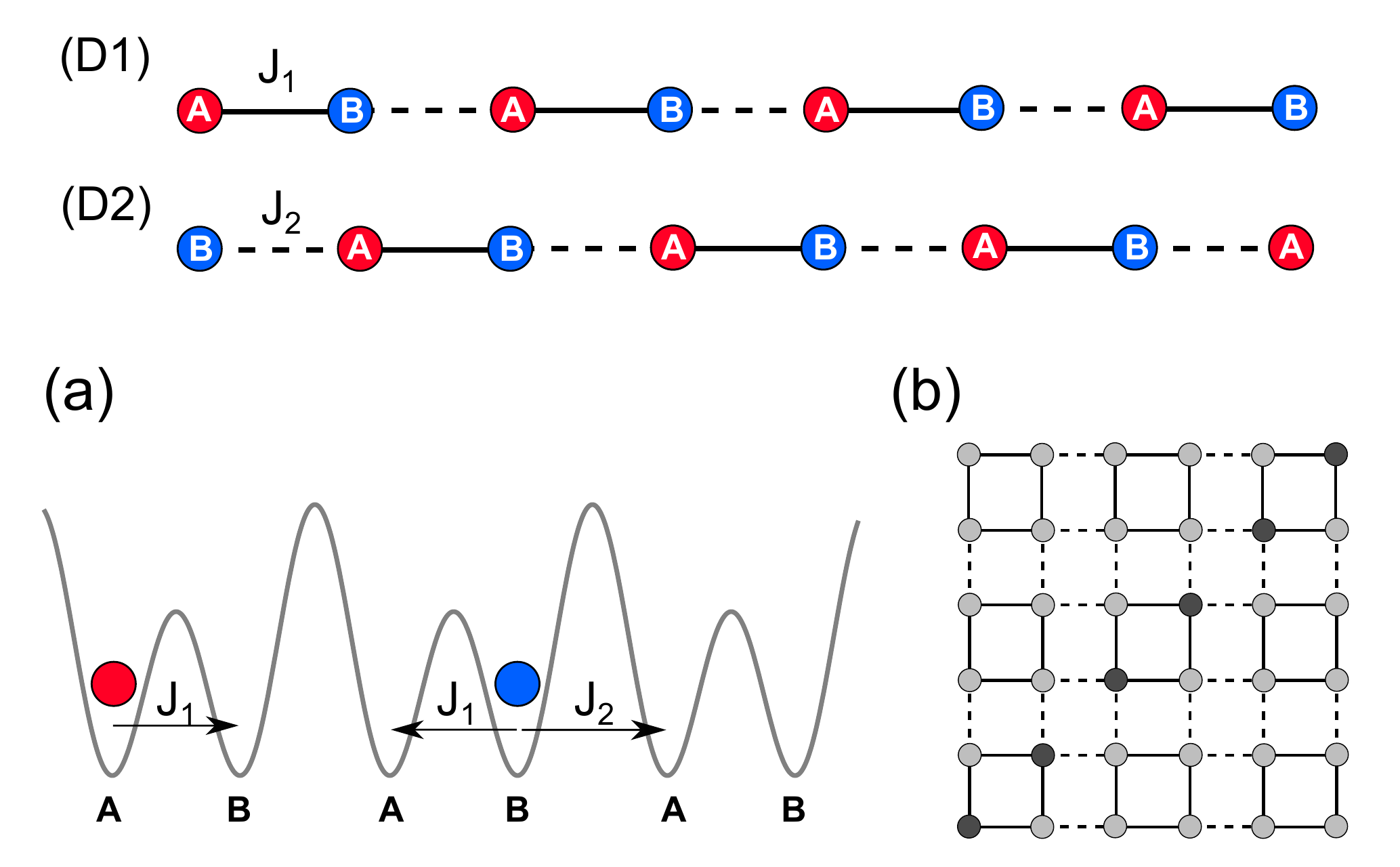}
\caption{Sketch of the SSH model considered in this work. For OBC and
  even number of sites, one can obtain two dimerizations: (D1)
  dimerization $D1$ starting and ending with a strong link $J_1$ and
  (D2) dimerization $D2$ starting and ending with a weak link $J_2$.
  (a) Example of two particles in a dimerized potential described by a
  Su-Schrieffer-Heeger model; (b) Sketch of the mapping onto a 2D
  single-particle system: strong links $J_1$ (full lines), weak links
  $J_2$ (dashed lines), and local potential $U$ (dark sites).}
\label{fig:lattice}
\end{figure}

In this work, we make an important step forward trying to combine a
topologically non-trivial single-particle band structure with
interactions. A prototypical phenomenon of this kind is the well
celebrated fractional quantum Hall effect
\cite{Tsui1982,Laughlin1983}.  Here, we focus our attention on the
minimal model of two interacting particles in a Su-Schrieffer-Heeger
(SSH) lattice. The full two-body spectrum can be calculated and very
rich physics emerges in spite of the simplicity of the model.  In
particular we find: (i) hybridization of different channels leading to
Fano-Feshbach resonances; (ii) existence of out-of-cell (long range)
bound pairs; (iii) edge states for the bound pairs. We conclude by proposing an experimentally
realistic optical fiber setup to quantum simulate the two-body SSH
model in the laboratory and experimentally highlight our predictions.

\begin{figure*}[!tbp]
\includegraphics[width=2.\columnwidth]{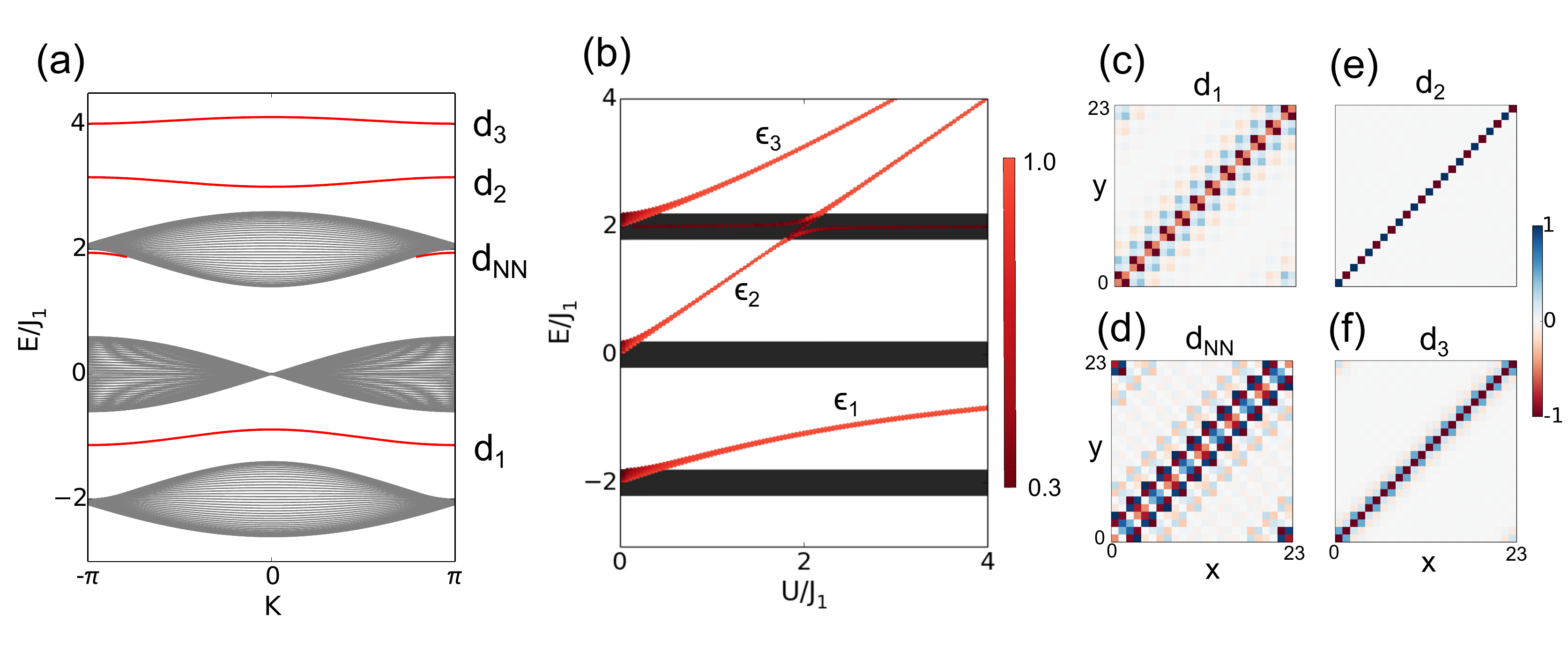}
\caption{(a) Two-body spectrum for PBC as a function of the center of
  mass momentum $K$ for $U=3 J_1$ and $J_2=0.3 J_1$, obtained from
  scattering theory. The spectrum presents three scattering continua
  and four bound states $d_{1}$, $d_{NN}$, $d_2$ and $d_3$. (b)
  Two-body spectrum as a function of the interaction $U$ for
  $J_2=0.1J_1$: The colorscale indicates the sum of onsite and
  nearest-neighboring site population and highlights the bound
  states. (c-f) Bound-state wave functions 
  for $U = 3 J_1$ and $J_2 = 0.1 J_1$ in a chain of $24$ sites at $K=0$ for $d_1$(c),
  $d_{2}$(e), $d_3$(f) and at $K=\pi$ for $d_{NN}$(d).}
\label{fig:PBC}
\end{figure*}

%%%% MODEL %%%% 

\section{Model}
\label{sect:model}

 In this work, we study two interacting bosonic particles
in the dimerized lattice shown in Fig.~\ref{fig:lattice} and governed
by the Hamiltonian $H = H_0 + H_U$, where
\be
\label{SSH}
H^{}_0 = - J_1 \sum_i c^\dag_{A,i} c^{}_{B,i} - J_2 \sum_i
c^\dag_{A,i+1} c^{}_{B,i} + \rm{H.c.}  
\ee 
is the kinetic part providing the single particle SSH model, whereas
\begin{equation}
\label{int_ham}
H^{}_U = \f{U}{2} \sum_i  \left( c^\dag_{A,i} c^\dag_{A,i} c^{}_{A,i} c^{}_{A,i} + c^\dag_{B,i} c^\dag_{B,i} c^{}_{B,i} c^{}_{B,i} \right)
\end{equation}
describes on-site interactions. For later convenience, we define a
lattice cell by a pair of $A$ and $B$ sites linked by tunneling
$J_1>J_2>0$, and label each lattice cell with index $i$. For periodic
boundary conditions (PBC) the two possible dimerizations are obtained
via a shift of a single lattice site, corresponding in practice to the
interchange of strong and weak tunneling. For open boundary conditions
(OBC) and even number of sites, we define $D1$ the dimerization
starting and ending with a strong link $J_1$ and $D2$ the dimerization
starting and ending with a weak link $J_2$ (see
Fig.~\ref{fig:lattice}(D1,D2)).

%%%% EXPERIMENTAL REALIZATION %%%%

In addition to ultra-cold atom implementations,
an interesting perspective of our work is 
to investigate the same physics with 2D lattices of side-coupled optical waveguides,
exploiting the mapping of two interacting particles in 1D
(Fig.~\ref{fig:lattice}(a)) onto a single particle in 2D
\cite{Longhi2011,Corrielli2013}. As sketched
in Fig.~\ref{fig:lattice}(b), the dimerized lattice is
reproduced by appropriately-tailored spatially-alternating hoppings
in the 2D lattice. Two-body on-site interactions in the 1D system
are translated into a local potential $U$ on the diagonal $x=y$ in
the single-particle 2D model. A straightforward extension of
existing experiments \cite{Schreiber2012,Mukherjee2015,Mukherjee2016},
would allow the possibility of observing distinctive two-body SSH dynamics 
directly in real space and real time.

%%%%%% BULK %%%%%%

\section{Bulk system}
 
 For periodic boundary conditions (PBC), the
single-particle SSH model possesses particle-hole symmetry
\cite{Ryu2002} and the spectrum is formed by two Bloch bands with
energy $E_{\pm}(k)=\pm \sqrt{J_1^2+J_2^2+2J_1J_2\cos(k)}$.  The two
possible dimerizations have the same spectrum, but present a Zak phase
difference of $\pi$ \cite{Zak1989,Atala2013}, corresponding to topologically
distinct phases identified by different winding numbers \cite{Ryu2016}.
Apart from the case of hard-core bosons at half-filling
(e.g. \cite{Grusdt2013,Deng2014}), the interacting Hamiltonian breaks
chiral symmetry and a typical two-body spectrum is shown in
Fig.~\ref{fig:PBC}(a) as a function of the center of mass momentum
$K$.

The essential spectrum is not modified by interactions and consists of
three scattering continua (type I), obtained by attributing an energy
belonging to the single-particle SSH spectrum to each scattering
particle. Instead, the wave functions of scattering states are modified by interactions, showing
a depletion at zero relative distance by increasing $U$.

In one dimension (1D), any interaction introduces a discrete spectrum,
related to the formation of bound pairs.  Three bound pairs are
readily identified by considering the fully dimerized case $J_2=0$.
For each cell $i$, the strong-link Hamiltonian admits the three
different states (written in the two-body basis $|A_iA_i\ket$,
$|A_iB_i\ket$ and $|B_iB_i\ket$)
\begin{align}
|d_{1,i}\ket &\propto \left( 2\sqrt{2} J_1, U + \sqrt{16J_1^2 + U^2} ,
2\sqrt{2} J_1 \right)\,, \label{d1}\\
|d_{2,i}\ket &\propto (1,0,-1)\,, \label{d2}\\
|d_{3,i}\ket &\propto \left( 2\sqrt{2} J_1, U - \sqrt{16J_1^2 + U^2} , 2\sqrt{2} J_1 \right)\,, \label{d3}
\end{align}
with energies $\epsilon_1=(U - \sqrt{16J_1^2 + U^2})/2$, $\epsilon_2 =
U$ and $ \epsilon_3 = (U + \sqrt{16J_1^2 + U^2})/2$.  \

For $J_2$ finite, the pairs delocalize along the lattice and develop narrow
bands (see Fig.~\ref{fig:PBC}).  The three bound states $d_\alpha$ can
be well defined for all values of $K$ at energies either in the band
gaps or above the continuum, or cross the continua in some parameter
range (see Fig.~\ref{fig:PBC}(b)).

Finally, at energies $\sim 2J_1$, an
additional {\it out-of-cell} bound state $d_{NN}$ appears,
characterized by a predominant contribution in neighboring cells
$(|A_i\ket - |B_i\ket ) \otimes ( |A_{i+1}\ket - |B_{i+1}\ket )$.
Such state arises thanks to an effective nearest-neighbor
interaction due to virtual processes involving mainly the $d_{2}$
state (see Appendix~\ref{app:dnn}). The $d_{NN}$ state
is present only for momenta around $K=\pi$. This fact can be understood
because the emergent nearest-neighbor interaction, which is responsible for the binding, 
is very weak compared to the bandwidth $2 J_2$ of the scattering continuum
(see for instance \cite{Valiente2009}). When $\epsilon_2=U\sim2J_1$, the
$d_{2}$ state becomes resonant with $d_{NN}$ and a strong
mixing between the two is observed Fig.~\ref{fig:PBC}(d).

%%%% SCATTERING THEORY %%%%

\subsection{Scattering theory}

The spectra of the bound states can be obtained by solving 
the Lippmann-Schwinger equation on the lattice.

For two particles, it is useful to describe the external degrees of
freedom using center-of-mass $R=(x+y)/2$ and the relative
coordinates $r=x-y$ for the two particles at lattice positions
$x$ and $y$, and the center-of-mass $K=k_1+k_2$ and relative
quasi-momentum $k=(k_1-k_2)/2$. As it usually happens for problems on
the lattice, the center-of-mass and relative coordinates do not
separate, but still, for PBC, the center of mass momentum $K$ is a
good quantum number, allowing to plot the spectrum as $E(K)$.  For the
sake of clarity, in a dimerized lattice of $N_c$ cells of lattice
spacing $D$ (corresponding to $N_s=2N_c$ lattice sites of lattice
spacing $d=D/2$) the allowed $K$ values in the first Brillouin zone
are given by $K=2\pi\ell/(N_cD)$ for $\ell=-N_c/2+1, \dots , N_c/2$, which,
upon Brillouin zone folding, coincide with the allowed $K$ values for
a uniform lattice $K=2\pi\ell/(N_sd)$ for $\ell=-N_s/2+1, \dots , N_s/2$.

To develop the scattering theory formalism, it is convenient to write
the SSH model in Eq.~(\ref{SSH}) in a different basis. After
performing the canonical transformation
\be
p^{}_i = \f{1}{\sqrt{2}} (c^{}_{A,i} + c^{}_{B,i})\,,\qquad m^{}_i = \f{1}{\sqrt{2}} (c^{}_{A,i} - c^{}_{B,i})\,,
\ee 
the single-particle Hamiltonian is cast into the form
\begin{align}
\label{SSHspin}
H'_0 =& -J_1 \sum_{i} \left( p^\dag_i p^{}_i - m^\dag_i m^{}_i \right) \\
- \f{J_2}{2}& \sum_{i,\, \nu=\pm1} \left( p^\dag_{i+\nu}p^{}_i - \nu p^\dag_{i+\nu} m^{}_i + \nu m^\dag_{i+\nu}p^{}_i - m^\dag_{i+\nu} m^{}_i \right)\,.\nn
\end{align}
Hamiltonian $H'_0$ describes a particle with pseudo-spin degrees
of freedom, labeled as $p,m$, hopping on a one-dimensional
lattice. This transformation is useful to treat the two-body problem
because the center of mass of each of the two single-particle states
$p,m$ is located in the middle of the $A-B$ bond. In a
first-quantization description, the two-body wavefunction can be
written as
\be |\Psi\ket = \sum_{x,y,\sigma}
\psi^\sigma(x,y) |x,y\ket \otimes |\sigma\ket \,, 
\ee 
where $x,y$ are, respectively, the unit-cell coordinates of
particle 1 and 2, and $|\sigma\ket \in \mathcal B_\sigma$ are the
two-body spin states
\be \mathcal B_\sigma=
\begin{cases}
|+\rangle = | p,p \rangle, \\
|0\rangle = \f{1}{\sqrt 2}\left( | p,m \rangle + | m,p \rangle \right),\\
|-\rangle = | m, m \rangle , \\
|F\rangle = \f{1}{\sqrt 2}\left( | p,m \rangle - | m,p \rangle \right).
\end{cases}\, \vspace*{0.5cm}
\ee
For the case of indistinguishable bosons discussed here, the amplitude
$\psi^\sigma(x,y)$ is symmetric when exchanging $x
\leftrightarrow y$, except for the $\sigma = F$ component, which 
must be antisymmetric in order to provide an overall symmetric wave function.
In the pseudo-spin basis, the interaction operator
$H^{}_U$ is still local, but not diagonal and represented by the
matrix
\be
H^{}_U= \frac{1}{2}
\begin{pmatrix}
 U & 0 &  U  & 0 \\
0 & 2U & 0 & 0 \\
 U & 0 &  U & 0 \\
0 & 0 & 0 & 0
\end{pmatrix}\,.
\ee
In the center-of-mass and relative coordinates, we make the standard ansatz
$\psi^\sigma(x,y) \equiv e^{i K R} \psi^\sigma(r)$ where
$K=k_1+k_2$ is the center-of-mass momentum. This choice of coordinates
and the choice of basis $\mathcal B_\sigma$ allow to decouple the
center of mass from the relative motion. After straightforward but
tedious calculations, we obtain the Schr\"odinger equation
\be
\left[ H^{2B}_0 + \delta_{r,0}H_U \right]_{\sigma,\sigma'} \psi^{\sigma'}(r) = E\, \psi^\sigma(r)\,,
\ee
where the kinetic part of the two-body Hamiltonian reads
\begin{widetext}
\be
H^{\textrm{2B}}_0=
\begin{pmatrix}
- 2J_1 - J_2 \cos(K/2) \Delta^+_r 	& -i \f{J_2}{\sqrt 2} \sin(K/2) \Delta^+_r & 0 &  \f{J_2}{\sqrt 2} \cos(K/2) \Delta^-_r\\
i \f{J_2}{\sqrt 2} \sin(K/2) \Delta^+_r  & 0 	 & -i \f{J_2}{\sqrt 2} \sin(K/2) \Delta^+_r & - i J_2 \sin(K/2) \Delta^-_r \\
0 &  i \f{J_2}{\sqrt 2} \sin (K/2)\Delta^+_r & 2 J_1 + J_2 \cos(K/2) \Delta^+_r &  \f{J_2}{\sqrt 2} \cos(K/2) \Delta^-_r\\
-\f{J_2}{\sqrt 2} \cos(K/2) \Delta^-_r 	& - i J_2 \sin(K/2) \Delta^-_r & -\f{J_2}{\sqrt 2} \cos(K/2) \Delta^-_r & 0
\end{pmatrix}\,.
\label{ham2Bscatt}
\ee
\end{widetext}
Here, we defined the discrete operators $\Delta^+_r \psi^\sigma(r) =
\psi^\sigma(r+1) + \psi^\sigma(r-1)$ and
$\Delta^-_r \psi^\sigma(r) = \psi^\sigma(r+1) - \psi^\sigma(r-1)$.

The Lippmann-Schwinger equation for the bound states reads

\begin{align}
\psi^\sigma (r) &=\mv{r\sigma | \hat G(E) \hat H_U | \psi} \nn\\
&= \sum_{\sigma',\, \sigma''}\int \f{dk}{2\pi} e^{ikr} G^{\sigma\sigma'}(k,E) (H_U)_{\sigma'\sigma''}\psi^{\sigma''}(0)\nn\\ 
&= \sum_{\sigma',\, \sigma''} G^{\sigma\sigma'}(r,E) (H_U)_{\sigma'\sigma''}\psi^{\sigma''}(0)\,,
\end{align}
where we have defined 
\be
\hat G(k,E) = (E - H^{2B}_0(K,k))^{-1}\,.
\ee
This formalism has been used to calculate the bound state spectrum
shown in Fig.~\ref{fig:PBC}(a).

%%%% RESONANT SCATTERING %%%%

\subsection{Resonant scattering}
\label{sect:resscatt}

The first noteworthy bulk feature of the two-body SSH model, persisting for
any boundary condition, is the strong mixing of the $d_2$ bound-state
narrow band and type I scattering continuum around the resonance
condition $U=2 J_1$, where the bound-state energy $U$ matches the
energy of a scattering state.  This mixing leads to a Fano-Feshbach
resonance in a lattice \cite{Nygaard2008, Valiente2009}, and can be
described analytically by using a two-channel scattering theory, as
shown below. The occurrence of scattering resonances due to repulsive
bound states in multi-band Hamiltonians has been studied also in other
contexts \cite{Valiente2011, Menotti2016}. 
  
\begin{figure}[!tbp]
  \includegraphics[width=1\columnwidth]{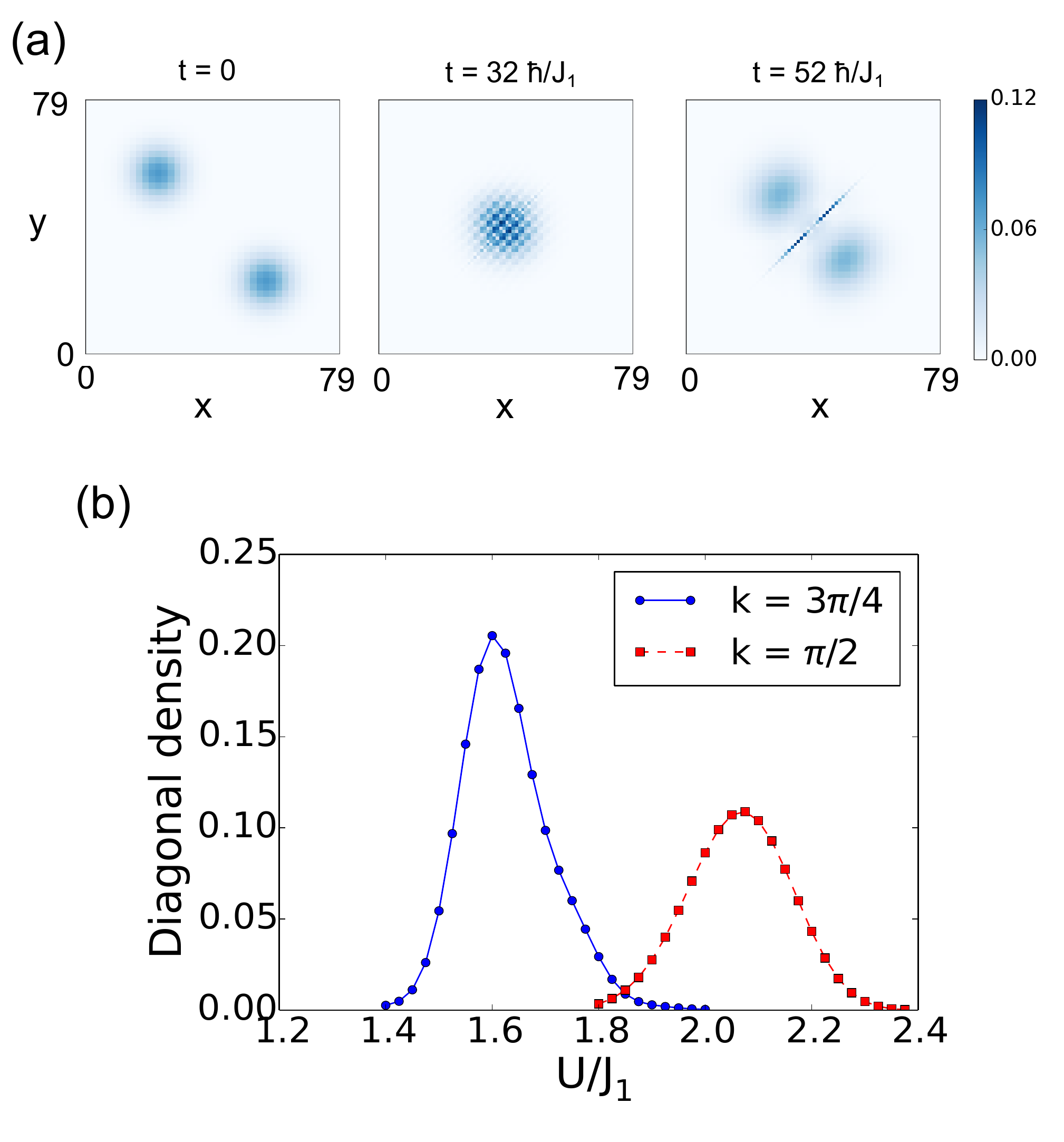}
  \caption{(a) Modulus of the two-body wavefunction
    $|\psi(x,y)|$ for two incident wave packets before, during and
    after collision for $J_2 = 0.1 J_1$, $U=2J_1$, $k=\pi/2$ and $L=80$ sites; time
    is measured in units of $\hbar/J_1$; (b) Diagonal density $\sum_x
|\psi(x,x)|^2$ after collision as a function of $U$ for two different $k$ and incident
    energies $E_{k=3\pi/4}=1.65 J_1$ and $E_{k=\pi/2}=2.08 J_1$.}
\label{feshbach}
\end{figure}

A Feshbach-like resonant scattering process is numerically illustrated
in Fig.~\ref{feshbach}, where we plot the square modulus of the 
two-body wave function $\psi(x,y)$ at times before, during and after the collision. 
At $t=0$ we prepare two single-particle gaussian wave packets
at momenta $k_1=k$ and $k_2=-k$ in the upper band of the SSH model,
localized at symmetric positions with respect to the lattice center,
sufficiently far from the boundaries and from each other, as shown
in Fig.~\ref{feshbach}(a). This initial state belongs to the two-body
scattering continuum centered around energy $2J_1$. The time evolution
in the presence of interactions $U$ is calculated numerically. After
collision, we observe two scattered wavepackets and a sizable
population of a two-body bound wavepacket of type $d_2$, highly
localized along $x=y$. At the beginning, the population of the bound
state is localized at the center of the lattice, then it expands at a
very slow rate along the $x=y$ direction while it decays in scattering
states.

In Fig.~\ref{feshbach}(b), we plot the diagonal density $\sum_x
|\psi(x,x)|^2$ providing a measure of the occupation of the bound
state at a time sufficiently after collision ($t=68\,\hbar/J_1$) for two
different values of the incident relative momenta, namely $k=3\pi/4$
and $k=\pi/2$, as a function of interaction $U$.  As expected, a clear
resonance peak is visible at $U$ such that the energy of the bound
state matches the energy of the incident wave packets. The different
heights of the two peaks can be understood as a consequence of the
finite life-time of the bound state and from the fact that the wave
packets are moving with different group velocities.

To obtain further understanding of these results,
one can perform a crude approximation and develop a theory including
only states $|0\ket$ and $|-\ket$. Indeed, $|0\ket \equiv
(|p,m\ket + |m,p\ket)/ \sqrt 2 = |A,A\ket - |B,B\ket$
is the dominant pseudo-spin component for $d_{2}$ 
when $J_2\ll J_1, U$. Analogously, the pseudo-spin state
$|-\ket = (|A\ket - |B\ket)\otimes (|A\ket - |B\ket)/2$ describes the
scattering states at energy $\sim 2J_1$.

\begin{figure}[!tbp]
\includegraphics[width=0.95\columnwidth]{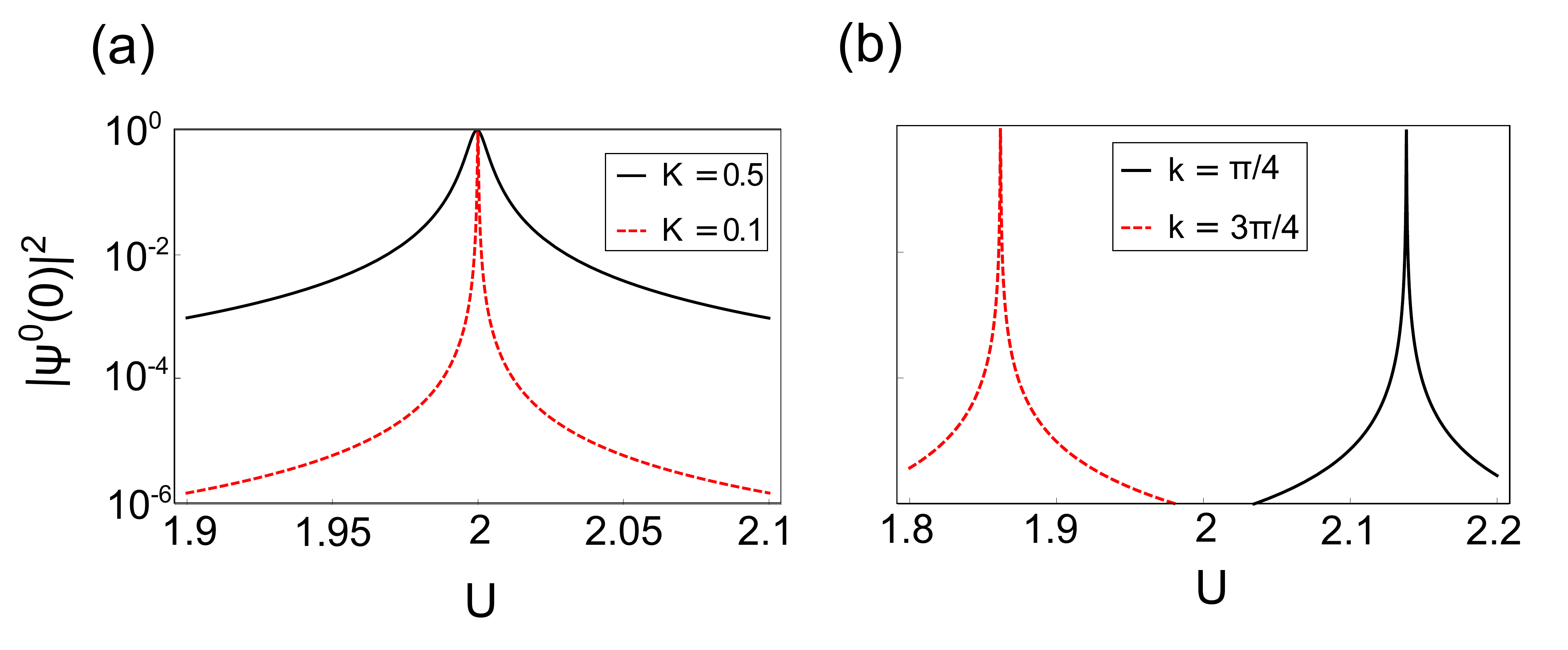}
\caption{Population $|\Psi^0(0)|^2$ in the $|0\ket$ component at $r=0$
  as a function of $U$ for: (a) center of mass momentum $K=0.5$ and
  $0.1$ at fixed $k=\pi/2$ and (b) relative momentum $k=\pi/2$ and
  $3\pi/4$ at fixed $K=0.3$. In all cases, the center of the resonance
  coincides with the energy $E$ of the scattering particles.}
\label{fig:peaks-scatt}
\end{figure}

The coupling of the other states $|+\ket$ and $|F\ket$ should be
introduced perturbatively in $J_2$.  However, since
Hamiltonian (\ref{ham2Bscatt}) already contains a coupling between
$|0\ket$ and $|-\ket$, the physics is captured in a qualitative manner
even neglecting all other states.
The reduced theory therefore reads
\be
\tilde H_0^\textrm{eff}(k) =
\begin{pmatrix}
0 	 & -i J_2 \sqrt 2 \sin\left(\frac{K}{2}\right) \cos k \\
i J_2 \sqrt 2 \sin\left(\frac{K}{2}\right) \cos k &
2 J_1 + 2J_2 \cos\left(\frac{K}{2}\right) \cos k 
\end{pmatrix} \nonumber
\ee
and 
\be
\tilde H_U = 
\begin{pmatrix}
U & 0 \\
0 & U/2
\end{pmatrix}.
\ee
The Green's function can be readily calculated and one finds
\begin{eqnarray}
  G_{11}(r) &=& \frac{ \delta_{r,0}}{E}, \\
  G_{22}(r) &=& i\f{e^{ikr}}{J_2 \cos(K/2) \sin k }, \nonumber \\
  G_{12}(r) &=& e^{ik r} \f{1}{\sqrt 2 E} \f{\tan(K/2)}{\tan k} + \delta_{r,0} \f{i}{\sqrt 2 E} \tan(K/2), \nonumber
\end{eqnarray}
where $E = 2J_1 + 2J_2 \cos(K/2)\cos k$ is the non-interacting
spectrum obtained neglecting the off-diagonal terms in $\tilde
H_0^{\textrm{eff}}$. The most general solution of the Schr\"odinger
equation is given by the two-component spinor $\Psi(r) =
(\Psi^0(r),\Psi^-(r))^T$:
\be \Psi(r) = \Phi(r) + G(r) \tilde H_U
(1-G(0) \tilde H_U)^{-1} \Phi(0)\,, \ee
where $\Phi(r)$ is solution of the non-interacting problem. To compare
with the numerical results presented above, we
take the ansatz $\Phi(r) = e^{ik r}(0,1)^T$.
According to this ansatz,
$\Phi(r)$ populates only the $|-\ket$ component and the two particles
have relative momentum $k$,
thus modeling two incident particles
belonging to type I continuum scattering off each other.

The population of bound state $d_2$ is described by the $|0\ket$
component of $\Psi(r)$ at $r=0$ (on-site pairs), namely
$\Psi^0(0)$. The results are shown in Fig.~\ref{fig:peaks-scatt}.  A
sharp resonance occurs at $U\sim E$, analogous to the one observed in
the numerical simulation of the dynamics of two colliding wave
packets. When the energy of the incident particles is close to $U$,
which is approximately the energy of the bound state, the probability
to form the bound state becomes maximal. Note how the resonance becomes sharper
when the center of mass momentum $K\ra 0$. Indeed, for $K=0$ the
off-diagonal terms in $\tilde H_0^\textrm{eff}$ vanish, thus
decoupling the two channels $|0\ket$ and $|-\ket$.

%%%%% EDGE STATES %%%%%%

\section{Edge physics}

\begin{figure*}[!tbp]
\includegraphics[width=2.\columnwidth]{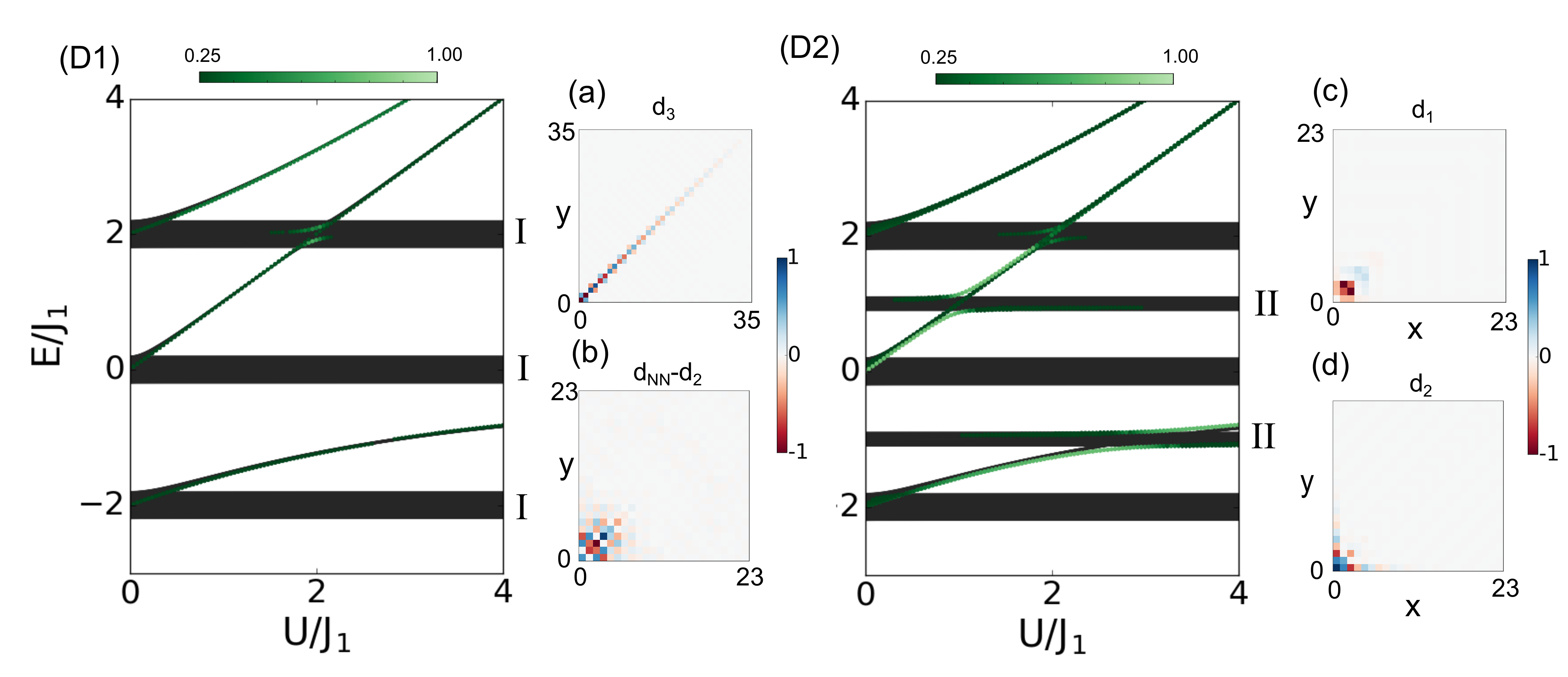}
\caption{Spectrum with OBC for $J_2=0.1 J_1$ as a function of $U$
  for (D1) dimerization $D1$ and for (D2) dimerization $D2$; The green colorscale
  represents the density in the first and last $2$ lattice cells and
  highlights the localization of EBS states; (a) $D1/d_3$ EBS for $U=3
  J_1$ and energy $E\sim 4 J_1$, (b) $D1/d_{NN}-d_2$ EBS for $U=2 J_1$ and energy $E\sim 1.9 J_1$, 
  (c) $D2/d_1$ EBS for $U=2 J_1$ and energy $E\sim-1.3 J_1$ and (d) $D2/d_2$ EBS for
  $U=J_1$ and energy $E\sim 0.8 J_1$ obtained by exact diagonalization of $72$ lattice sites.}
\label{spectrum_OBC_U}
\end{figure*}

We now discuss the case of open boundary conditions (OBC) to address
the effect of interactions on the finite chain SSH model. As usual, we
need to distinguish the two possible dimerizations $D1$ and $D2$. Single-particle 
edge states, typical of dimerization $D2$, combined with a
freely propagating particle generate two further continua around
energies $\pm J_1$ (type II). Obviously, such type II continua are
absent in $D1$, which does not admit single-particle edge states (see
Fig.~\ref{spectrum_OBC_U}(D1,D2)). The two (type I and type II)
continua and the narrow bands of bound states are independent
consequences of the single-particle SSH model and of two-body
interactions, respectively.  Instead, as a pure consequence of the
interplay between SSH geometry, interactions and boundary conditions,
intriguing two-body edge bound states (EBS) emerge in the spectrum
(see Fig.~\ref{spectrum_OBC_U}(a-d)).
Their presence or absence is highly non-trivial and essentially driven by a
renormalization of the edge properties (see also 
Refs.~\cite{Haque2009,Platero2016}) .

Such EBS can be associated to the different bound states $d_i$.  For
PBC the associated bound states - when well defined in the whole
Brillouin zone - present a two-particle generalized \footnote{See, for instance, Ref.~\cite{Lee2016} 
for a generalization of topological invariants to multiparticle systems.} 
Zak phase difference of $\pi$ for the two
dimerizations $D1$ and $D2$. However, as we are going to show in the
following, this does not necessarily correspond to the formation of
EBS in the finite chain, leading to a breaking of the standard
bulk-boundary correspondence.  Furthermore, most remarkably, as
clearly visible in Fig.~\ref{spectrum_OBC_U}(D1,D2), EBS appear not
only in dimerization $D2$ but also in dimerization $D1$, which does
not admit edge states in the single-particle case.

The $D1/d_3$ and $D2/d_{1,2}$ EBS can be interpreted as Tamm states of
an effective strong-dimerization theory, as it will be detailed
below. Localization persists even when the EBS energy enters the
scattering continua for $U\ra 0$ \cite{Soljacic2016, Longhi2014}.  Moreover, immersed in the higher
type I scattering continuum, we find a further EBS, which is present
in both dimerizations and can be related to the existence of the
out-of-cell bound state $d_{NN}$. Hybridization between 
$d_{NN}$ and $d_2$ around $U\sim 2J_1$ induces a very strong localization at the
edges of a wavefunction with strong both diagonal and out-of-cell
characters (see Fig.~\ref{spectrum_OBC_U}(b)).

%%%%%%%%%%%%%%%%%%%%%%
\subsection{Strong-dimerization limit}
%%%%%%%%%%%%%%%%%%%%%%

In order to understand the physics behind bound states and EBS, it is
useful to consider the regime of strong dimerization $J_2 \ll J_1,
U$. Here, effective models accounting for the weak tunneling $J_2$ in
second order perturbation theory can be developed. The building blocks
for the effective theory are naturally the three strong-link two-body
eigenstates given in Eqs.~(\ref{d1}-\ref{d3}). The effective lattice
is provided by the lattice cells $i$. More details can be
found in Appendix~\ref{eff}.

In-cell dimers $d_{\alpha,i}$ can tunnel at second order through
intermediate states given by a particle in link $i$ and a particle in
a neighboring link $j$. The effective model reads
\be
\label{effHam2}
H_{\textrm{eff}} = \sum_{i,\alpha} E_{\alpha,i}\, d^\dag_{\alpha,i} d_{\alpha,i} + \sum_{\mv{i,j}}\sum_{\alpha,\beta} J_{\alpha\beta}^{ij}\, d^\dag_{\alpha,i} d_{\beta,j}\,.
\ee
The parameters that appear in the model above are second order in the
weak tunneling $J_2$.  The effective model in Eq.~(\ref{effHam2})
provides an accurate prediction of the bound state spectrum away from
$U\sim J_1$ and $U\sim 2J_1$ where bound state $d_2$ crosses type II
and type I scattering continua, respectively, or $U \sim 3 J_1$ where
$d_1$ crosses the lower type II continuum. Relying on the additional assumption
that the bound states are well separated in energy and the coupling
among them is weak, effective model (\ref{effHam2}) can be further
simplified through a single band approximation, which only keeps
$J_{\alpha\alpha}^{ij}$ and $E_{\alpha,i}$ for each state
$d_{\alpha,i}$.

\begin{figure}[!tbp]
  \includegraphics[width=0.9\columnwidth]{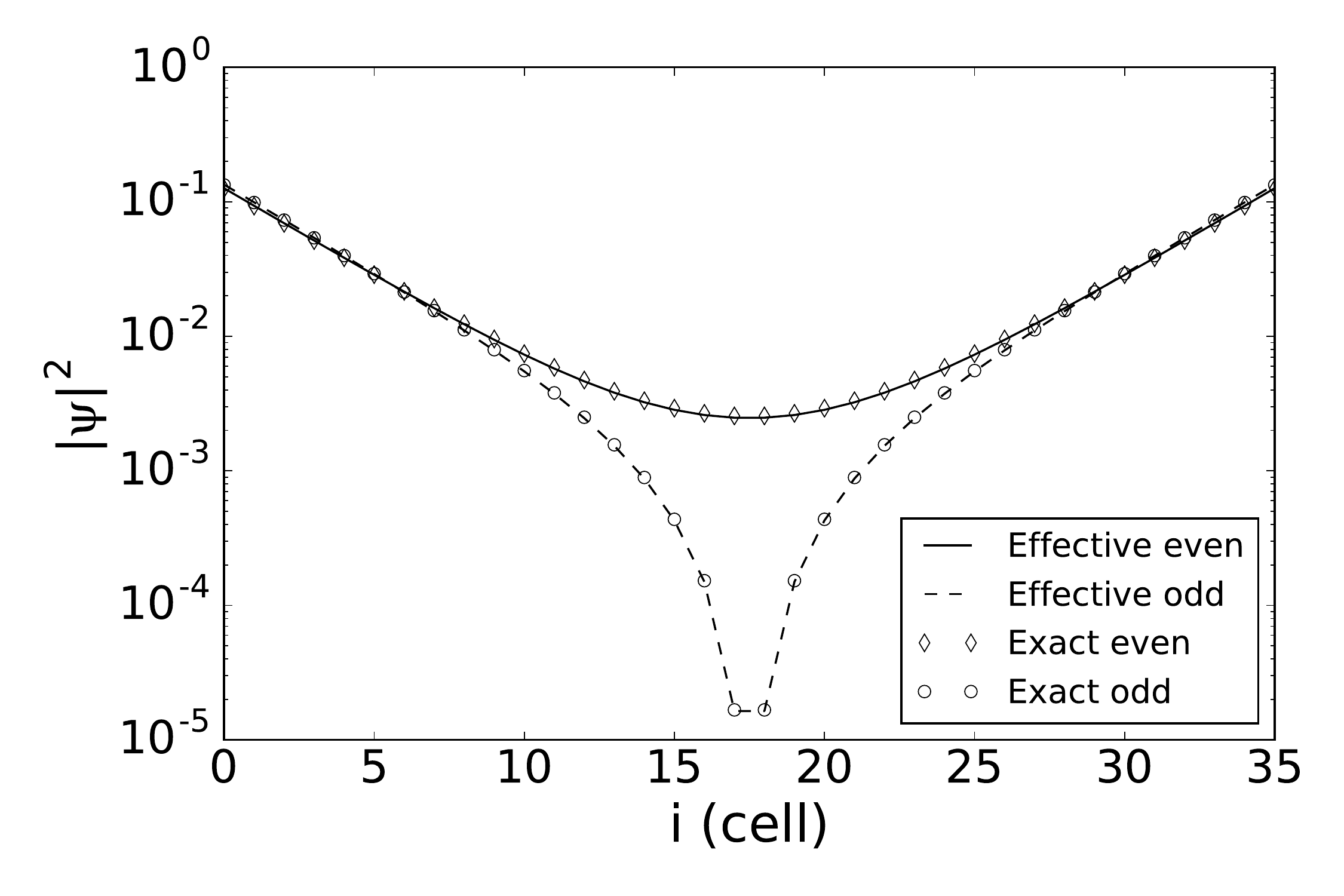}
  \caption{Exponential localization of the $D1/d_3$ edge state for $U=3J_1$ and $J_2=0.1 J_1$ 
  in a lattice with $36$ unit cells ($72$ sites) in dimerization $D1$. We
  plot the probability to find two particles in unit cell $i$  
  calculated with the effective theory in the strong dimerization limit (lines) 
  and with exact diagonalization (markers). Both simulations provide a pair of
   even and odd almost-degenerate eigenstates due to finite size (see legend).}
\label{fig:D1d3Comp}
\end{figure}

Just below the $d_3$ bound-state narrow band in dimerization $D1$, one
finds EBS $D1/d_3$ (see Fig.~\ref{spectrum_OBC_U}(a)), which can be
quantitatively explained as a Tamm state in the framework of effective
model (\ref{effHam2}).  The comparison between the results
obtained with exact diagonalization and with the effective model is shown in
Fig.~\ref{fig:D1d3Comp}.  The localization length of EBS
$D1/d_3$ is very large. It increases for strong interactions $U \gg
J_1,J_2$, so that, for practical purposes, in a finite lattice this
state undergoes a crossover to a not exponentially-localized
state. 

A deeper understanding of the physics underlying the divergence
of the localization length  for large interactions can be obtained via a much simpler {\it
  strong-interaction effective model.} The states $d_2$ and $d_3$ 
  are almost degenerate for $U\gg J_1, J_2$. In this limit, 
  a more convenient basis is given by on-site doublons
$d^\dag_{A,i}|0 \ket\equiv|A_iA_i\ket$ and $d^\dag_{B,i}|0
\ket\equiv|B_iB_i\ket$ coupled among each other via second order
processes.  The corresponding effective Hamiltonian is nothing else
than an effective single-particle SSH model with effective hopping coefficients
$J_{1,2}^\textrm{eff}=-2J_{1,2}^2/U$ and effective on-site energy
$\epsilon_{\textrm{bulk}}= U + 2 ( J_1^2 + J_2^2)/U$.  Moreover, for
dimerization $D\sigma$ (with $\sigma=1,2$), the on-site energy of a
doublon at the edge results $\epsilon_{\textrm{edge}} =
\epsilon_{\textrm{bulk}} +\Delta E_\sigma$, with $\Delta E_\sigma =-
2J_{3-\sigma}^2/U$.  This energy shift at the outermost sites provides
a generalization of the Tamm physics to the SSH model, which in
general, depending on $\Delta E$, allows both for Tamm-like states
above or below the continua and in-gap states. However, the specific
case of our effective model coincides, in both dimerizations, exactly
with the critical value of $\Delta E$ for which neither Tamm nor
in-gap states can exist (see discussion in Appendix~\ref{app_strong_int1}). 
This implies that in the strong-interaction
limit $U\gg J_1$ exponentially localized edge states are not to be expected
in finite-size chains, in agreement with the numerical results.

For dimerization $D2$, a closer inspection of the $d_1$ and $d_2$
dimer spectra around their intersection with type II continua shows a
peculiar feature: two dimer states, gapped from their continua, appear
(see Fig.~\ref{spectrum_OBC_U}(c-d)).
They correspond to pairs of $D2/d_{1,2}$ EBS moved out of the corresponding
bound-state narrow bands as a consequence of the renormalized
parameters at the boundaries. 

\begin{figure}[!tbp]
  \includegraphics[width=0.95\columnwidth]{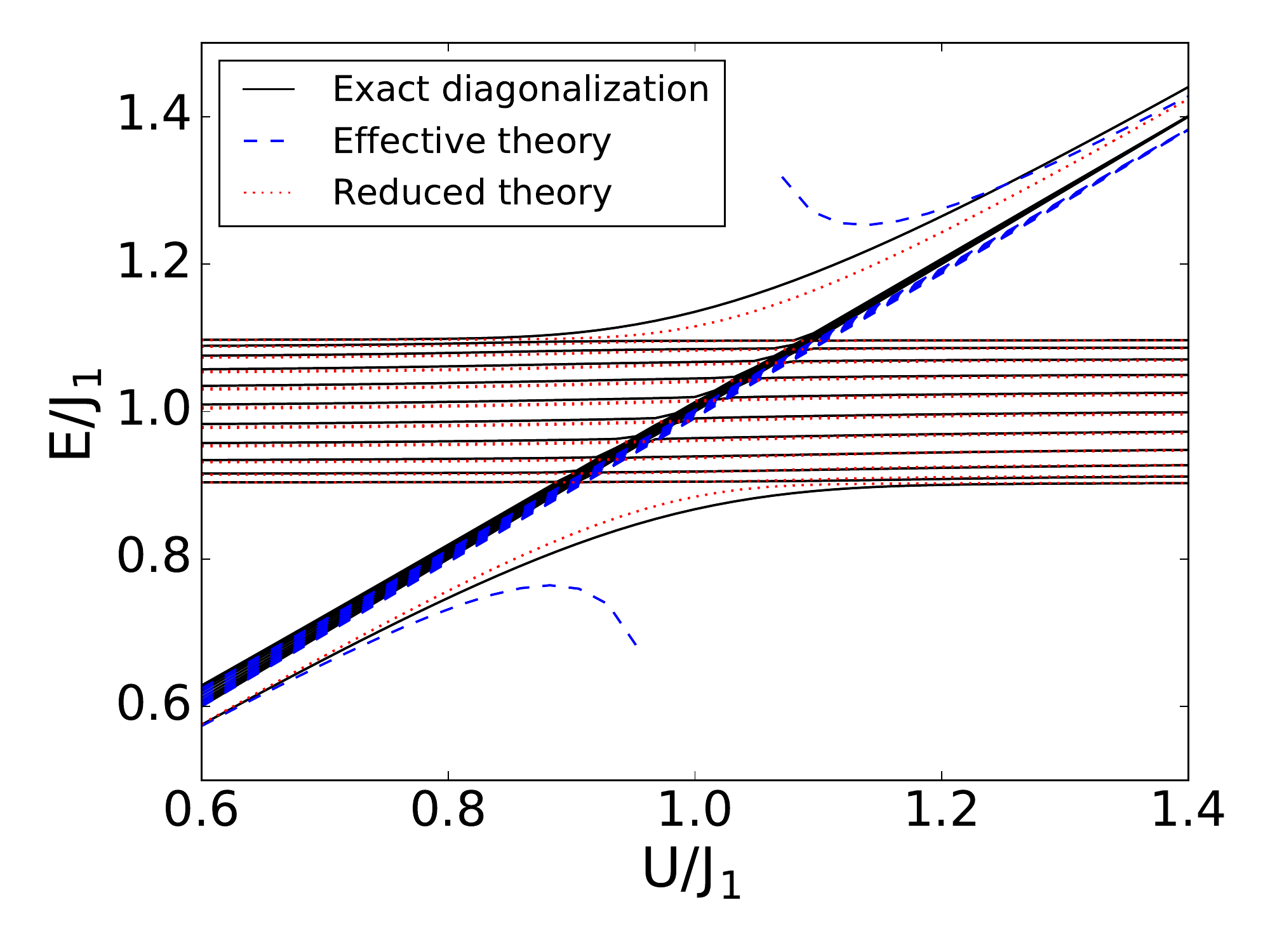}
  \caption{Energy spectrum in dimerization
    $D2$ at the crossing between the $d_2$ bound state and the upper
    type II continuum for $J_2=0.1 J_1$ and $24$ lattice sites: Exact
    diagonalization results (black lines), strong dimerization model
    (blue dashed lines) and reduced model (red dotted lines).}
\label{fig:crossing}
\end{figure}

These EBS can also be accounted for by effective model (\ref{effHam2}).
In $D2$, the effective model is slightly
more complicated than in $D1$, since no full lattice cell is present
at the edges, but rather a single lattice site (see
Fig.~\ref{fig:lattice}(D2)). In the following, we thus specialize to the 
case of the $D2/d_2$ state. While in the bulk the bound state
preserves its standard form $|d_{2,i}\ket = (|A_i,A_i\ket -
|B_i,B_i\ket)/\sqrt 2$, for the doublons at the edges one needs to
consider the ansatz $|d_{2,0}\ket = -|B_0, B_0\ket$ and $|d_{2,L}\ket
= |A_L, A_L\ket$. This truncated bound-state wave function affects
both the effective hopping $J_{22}^\textrm{edge,D2} \neq
J_{22}^\textrm{bulk}$, the on-site energy $E_{2}^\textrm{edge,D2} \neq
E_{2}^\textrm{bulk}$ at the edges, and the on-site energies
$E_{2,1},E_{2,L-1}$ at the outermost complete lattice cells.  As shown
in Fig.~\ref{fig:crossing} (blue lines), sufficiently far away from
the $U=J_1$ condition, the effective model perfectly reproduces the
numerical spectrum and the presence of gapped states.

The effective theory fails at $U \sim J_1$ because the $D2/d_2$ EBS
becomes resonant with type II scattering states. In order to account 
for the hybridization mechanism, we consider a reduced Hilbert space including $d_2$-like
truncated EBS $-|B_0, B_0\ket$ and $|A_L, A_L\ket$, type II scattering
states $|\psi^{j}_{i}\ket = |ES_j\ket \otimes \left(|A_i\ket -
|B_i\ket \right)/\sqrt 2$, where $|ES_0\ket=|B_0\ket$ and
$|ES_L\ket=|A_L\ket$ and finally the zero-energy single-particle edge 
state $|B_0 A_L\ket$ (see Appendix~\ref{app:reduced}).  This reduced theory
reproduces very well the avoided crossing around $U=J_1$ (see
Fig.~\ref{fig:crossing} (red dotted lines)) and points out
  that the $D2/d_{2}$ EBS is smoothly transformed into a type II
  scattering state when moving away from $U\sim J_1$.

%%%%%%% DYNAMICS %%%%%%

%%%%%%%%%%%%%%%%%%%%%%%%%%%%%%%%
\section{Experimental observation and dynamics}
%%%%%%%%%%%%%%%%%%%%%%%%%%%%%%%%
\label{exp}

  In this final section, we present the results of real time simulations which
  directly highlight the properties of the SSH model discussed in this
  work. The most promising idea is that, upon the 1D to 2D mapping
  introduced in Sec.~\ref{sect:model}, a 2D coupled optical fibers
  setup can provide a quantum simulator of the two-particle 1D SSH
  model, such that the full two-body dynamics of the system is visibile in real
  time and real space through the propagating light intensity.
Beyond the characterization of scattering, bound and edge bound states
as discussed in this section, the same setup would allow the
visualization of the closed channel population in a
resonant Fano-Feshbach scattering process, already presented in
Sec.~\ref{sect:resscatt}.

We study the two-body dynamics, assuming different initial conditions
at $t=0$ and different interaction strengths $U$. We let the two-body
wave function in second quantization evolve numerically in time via
exact diagonalization. Written in first quantization, the two-body
wavefunction can be interpreted as a single particle wave
function $\psi(x,y,t)$ in 2D ($x$ and $y$ being
  equivalently the coordinates of the two particles in 1D or the
  coordinates of a single particle in 2D). We address few different 
  illustrative cases, shown in the following subsections.

%%%%%%%%%%%%%%%%
\subsection{Edge bound state $D2/d_2$}
\label{ebs_exp}
%%%%%%%%%%%%%%%%

\begin{figure}[!tbp]
\includegraphics[width=1\columnwidth]{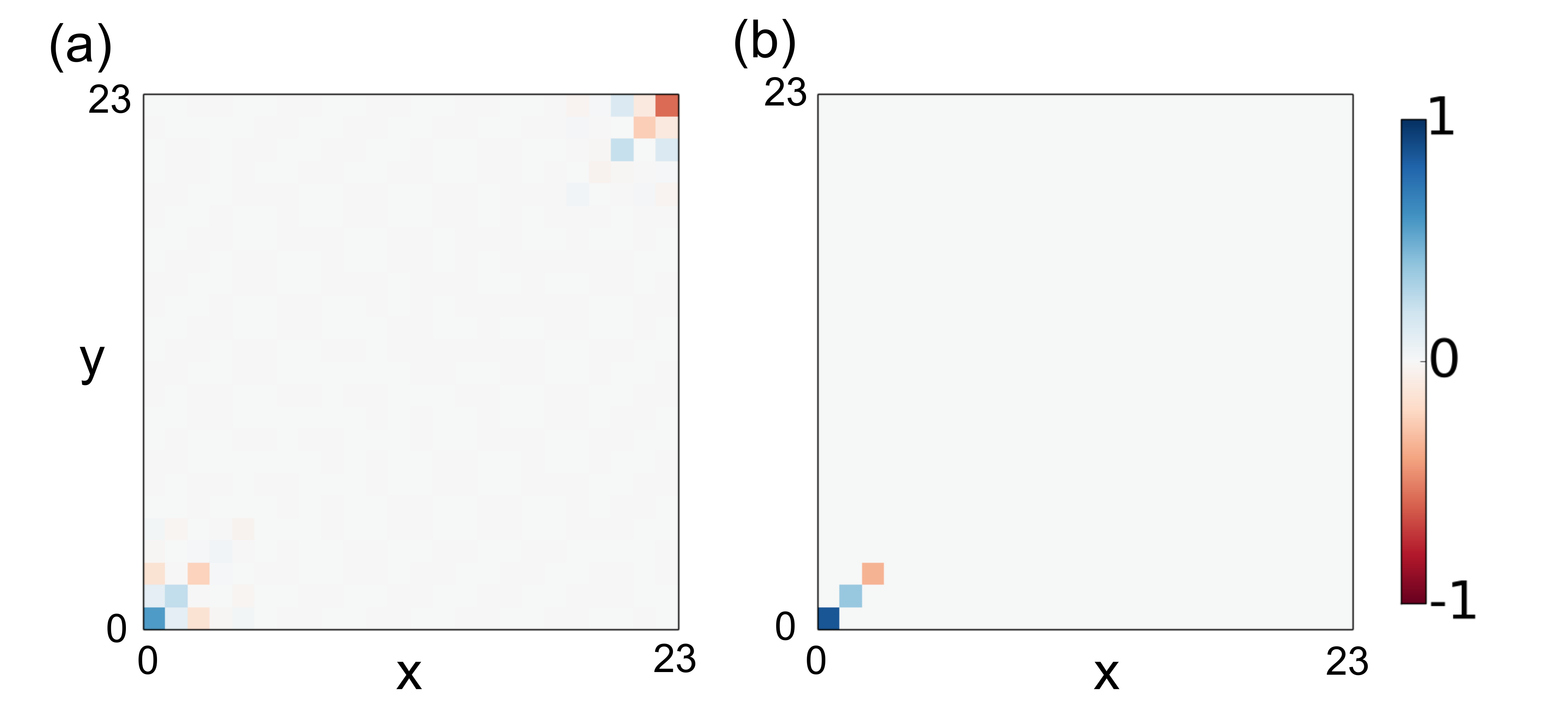}
\caption{(a) EBS wave function for a lattice with $L=24$ sites in the
  D2 dimerization for $J_2=0.1 J_1$, $U=0.7 J_1$ obtained with exact
  diagonalization. (b) Projected wave function on $x=y$ with $x \leq
  3$, used as initial state for the time-evolution shown in Fig.~\ref{fig:tEvol-loc}(a)
  and discussed in Sec.~\ref{ebs_exp}.}
\label{fig:edge_init}
\end{figure}

In Fig.~\ref{fig:edge_init}(a), we plot the wave function of the exact
EBS eigenstate for $U = 0.7 J_1$ and $J_2 = 0.1 J_1$ in dimerization
$D2$ (see Fig.~\ref{fig:crossing}). The main components of the EBS wave function are on the diagonal
$x=y$ and decay exponentially as $x$ grows. Being even
  and odd states almost degenerate, one can equally well consider
  states localized at either end of the lattice. Therefore, as initial
  state, we take the projection of the exact EBS wave function on
  $x=y$ with $x \leq 3$, as shown in Fig.~\ref{fig:edge_init}(b),
  localized at the bottom left corner of the 2D lattice.

\begin{figure}[!tbp]
\includegraphics[width=1\columnwidth]{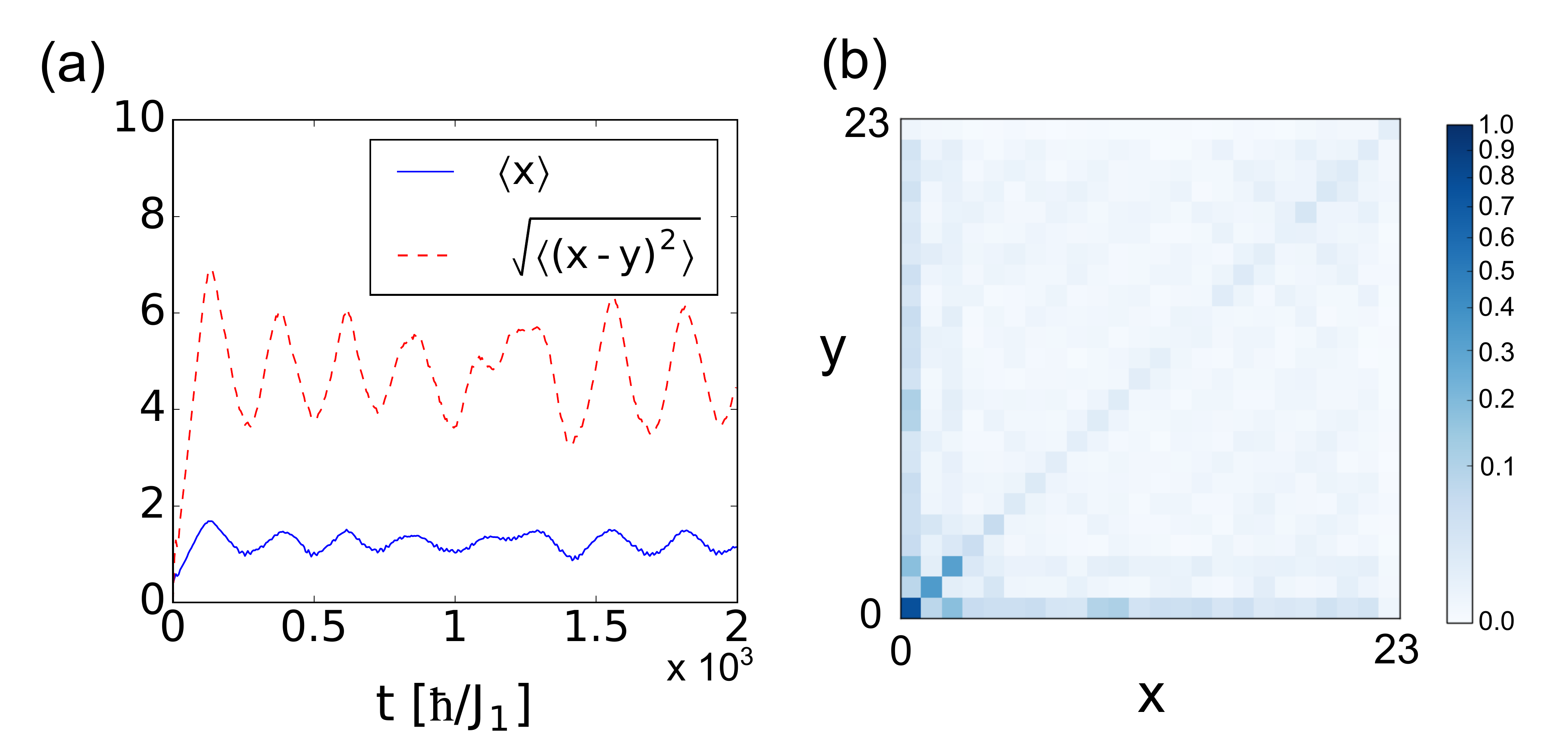}
\caption{(a) Time evolution of the projected EBS in
  Fig.~\ref{fig:edge_init}(b) for dimerization $D2$: $\langle x
  \rangle$ (full blue line) and $\sqrt{\langle (x-y)^2\rangle}$ (dashed red line) as a function of time; 
  (b) Modulus of the two-body wavefunction at $t=
  10^3 \hbar/J_1$. In these simulations $U=0.7J_1$, $J_2=0.1J_1$ and $L=24$ sites.}
\label{fig:tEvol-loc}
\end{figure}

We use the observables $\mv{\hat x}(t) \equiv \sum_x x\, |\psi(x,y,t)|^2 =
\sum_y y\, |\psi(x,y,t)|^2$ and $\sqrt{\mv{(\hat x - \hat y)^2}}(t)$
to characterize the edge
localization properties of the states.  In Fig.~\ref{fig:tEvol-loc}(a)
the time evolution of $\mv{\hat x}(t)$ is displayed. The plot shows
that $\mv{\hat x}(t) \ll L/2$, namely the initially approximate EBS,
remains localized at one edge of the system. It is remarkable that
a very well approximated EBS can be obtained  by initializing the
wave function over only three lattice sites.

However, since the
initial state slightly differs from the exact EBS, a small
overlap with type II states is present and observed in a 
non-vanishing single-particle population oscillating
at $x=0$ or $y=0$ (see Fig.~\ref{fig:tEvol-loc}(b)). This produces sizable -
but still small when compared to the lattice size -
fluctuations $\sqrt{\langle( x - y)^2\rangle}$. The visible oscillations in both observables
arise from the bouncing of the populated type II states at the lattice edges.

\begin{figure}[!tbp]
\includegraphics[width=1\columnwidth]{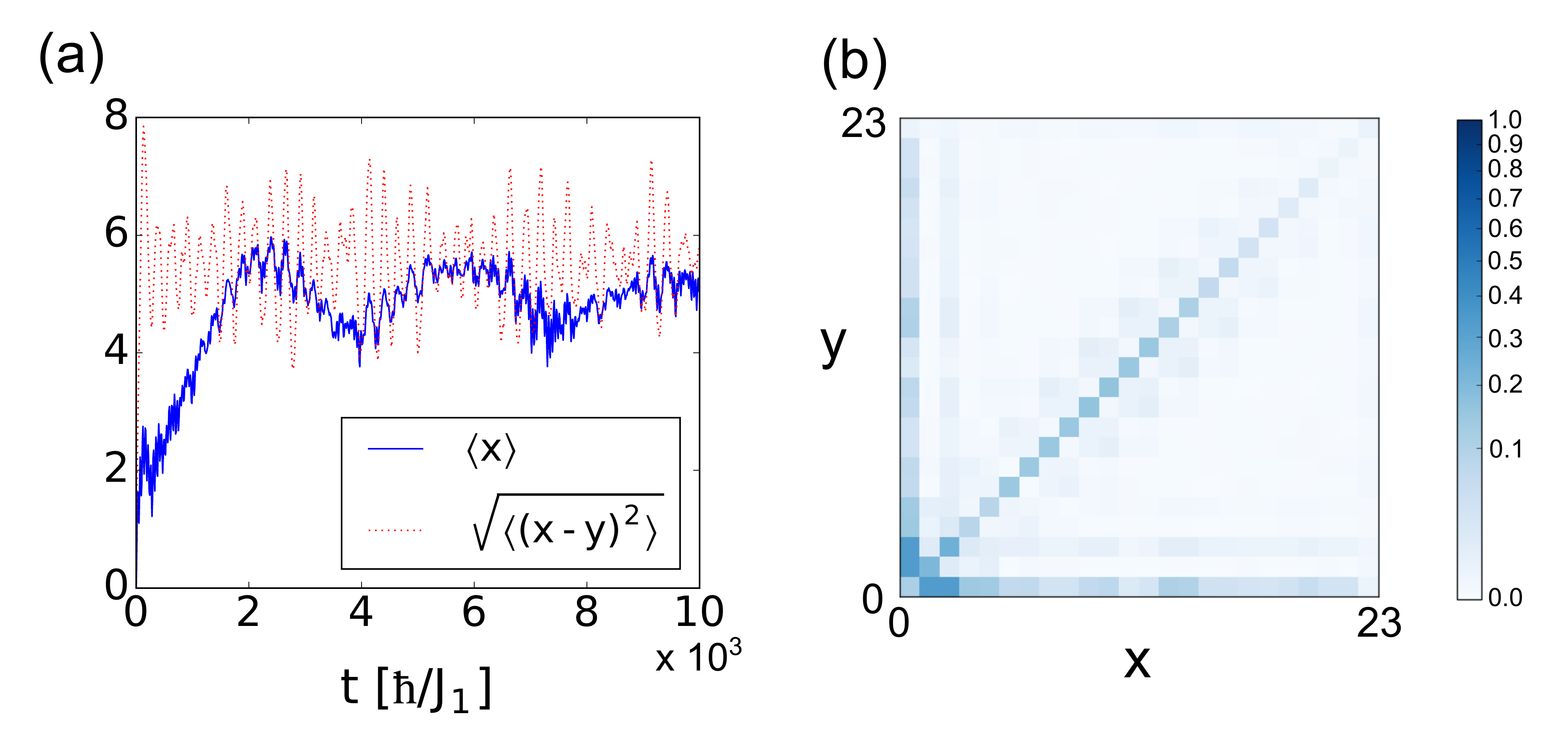}
\caption{(a) Time evolution of a doublon initially localized at the outermost lattice 
  site for dimerization $D2$: $\langle x
  \rangle$ (full blue line) and $\sqrt{\langle (x-y)^2\rangle}$ (dotted red line) as a function of time;
  (b) Modulus of the two-body wavefunction at $t=
  10^3 \hbar/J_1$. In these simulations $U=J_1$, $J_2=0.1J_1$ and $L=24$ sites.}
\label{fig:tEvol-deloc}
\end{figure}

\begin{figure}[!tbp]
\includegraphics[width=1\columnwidth]{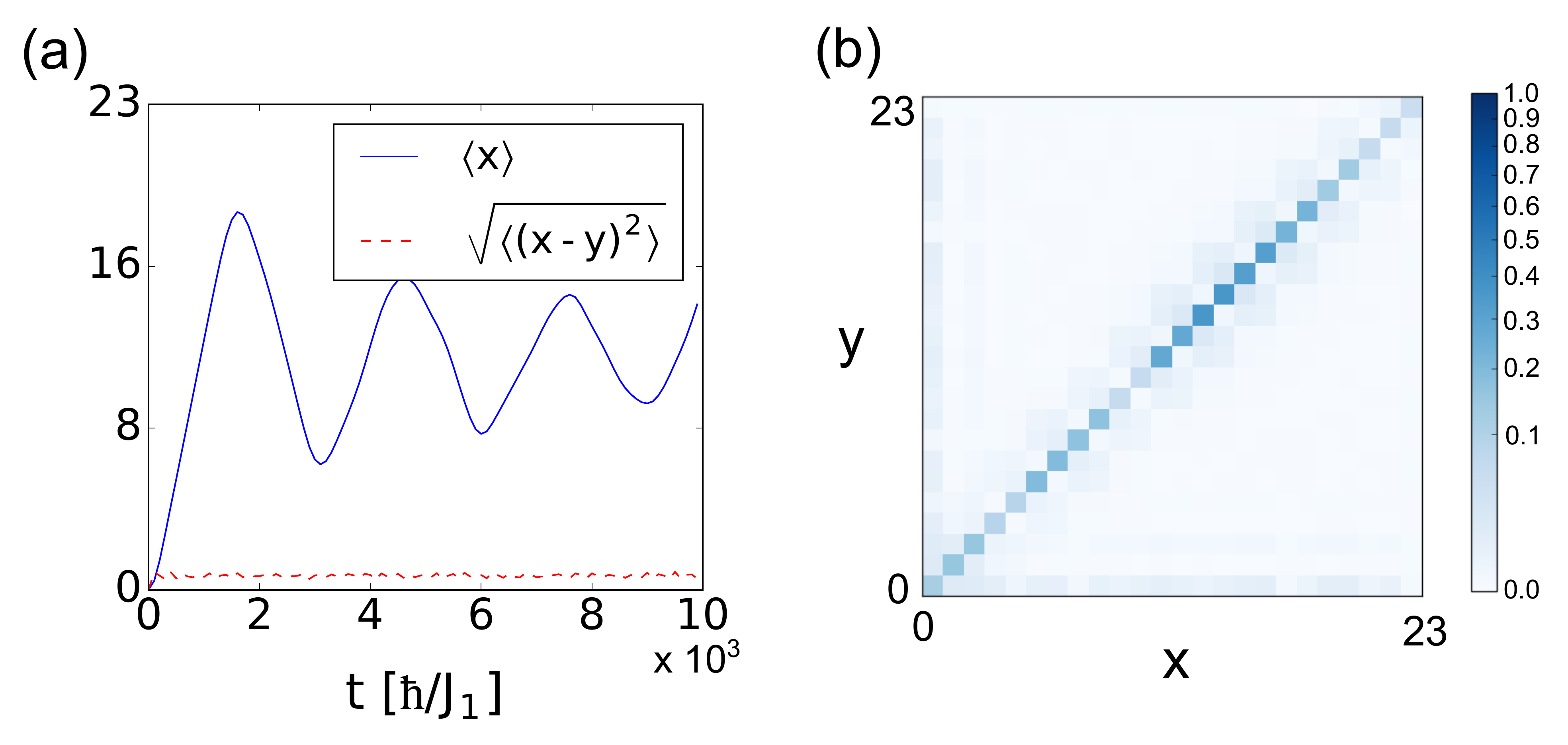}
\caption{(a) Time evolution of a doublon initially localized in the
  lattice outer most lattice site for dimerization $D2$: $\langle x
  \rangle$ (full blue line) and $\sqrt{\langle (x-y)^2\rangle}$ (dashed red line) as a function of time;
  (b) Modulus of the two-body wavefunction at $t=
  10^3 \hbar/J_1$. In these simulations $U=3J_1$, $J_2=0.1J_1$ and $L=24$ sites.}
\label{fig:tEvol-d2}
\end{figure}

%%%%%%%%%%%%%%%%%
\subsection{Hybridization between $d_2$ and type II continuum}
%%%%%%%%%%%%%%%%%

Differently from the previous section, we consider as initial 
condition a single doublon localized at the 
outermost site of a $D2$ dimerized lattice, and tune the
value of interactions to $U=J_1$. 
Such initial state has a sizable overlap with the $D2/d_2$ EBS, 
the $d_2$ bound-state continuum and type II scattering states. 

The time evolution shows that the state again remains mostly localized
at one edge. However, $\mv{\hat x}(t)$ becomes larger because of the non-negligible 
population of the $d_2$ continuum (see Fig.~\ref{fig:tEvol-deloc}(a)). 
Moreover, oscillations at two different characteristic time-scales are visible. 
The fast time scale is present in both observables 
and it is related to the bouncing of the type II
states, as discussed in the previous section. A second slower time-scale is clearly recognizable for the
observable $\mv{\hat x}(t)$ related to the dynamics of the heavy $d_2$ bound state and the 
corresponding bouncing off the lattice edges. 

%%%%%%%%%%%%%%%%
\subsection{Bound state dynamics ($D2$)}
%%%%%%%%%%%%%%%%

 As initial state, we take again a single
doublon localized in the outermost site of a $D2$ dimerized lattice,
but increase interactions to move away from the resonance between
$d_2$ and type II continuum.

At $U=3J_1$, this initial state has a large overlap
with the $d_2$ bound states, which are well localized at
$x=y$, but not necessarily at the edges, 
and negligible overlap with the scattering continua.  For that reason, the state delocalizes
in the lattice remaining bound at relative distance equal to zero, as
shown in Fig.~\ref{fig:tEvol-d2}.
This is reflected in a negligible value of $
\sqrt{\langle(x-y)^2\rangle}$ during the whole time evolution and a
center-of-mass average position of the wave packet oscillating
significantly in time due to bounces at the lattice edges.

%%%%%%%%%%%%%%%%
\subsection{Two-body scattering states}
%%%%%%%%%%%%%%%%

As a final example, we show the case where we
populate and address scattering states.

We take as initial condition a state delocalized in the first four
lattice sites cells without any double occupation. In our notations,
the initial state reads differently in the two dimerizations, so that
it is convenient to write it explicitly (symmetrization is assumed):

\begin{eqnarray}
 |\Psi_{D1}(t=0)\rangle& \propto& (|A_1\rangle+|B_1\rangle) \otimes
 (|A_2\rangle+|B_2\rangle), \label{but1}\\ |\Psi_{D2}(t=0)\rangle&\propto&
 (|B_0\rangle+|A_1\rangle) \otimes
 (|B_1\rangle+|A_2\rangle). \label{but2}
\end{eqnarray}
Due to the vanishing double occupation, the energy of the initial
state is determined by the hopping processes and not by
interactions. It results $E^{(in)}_\sigma=-2J_\sigma$ depending on the
dimerization $D\sigma$ ($\sigma=1,2$). In dimerization $D1$ the
  initial state lies almost completely in the lower type I scattering
  continuum, having very small projection on other states. Instead, in
  dimerzation $D2$, the initial state has non negligible overlap with states in
  all type I and type II continua.

\begin{figure}[!tbp]
\includegraphics[width=1\columnwidth]{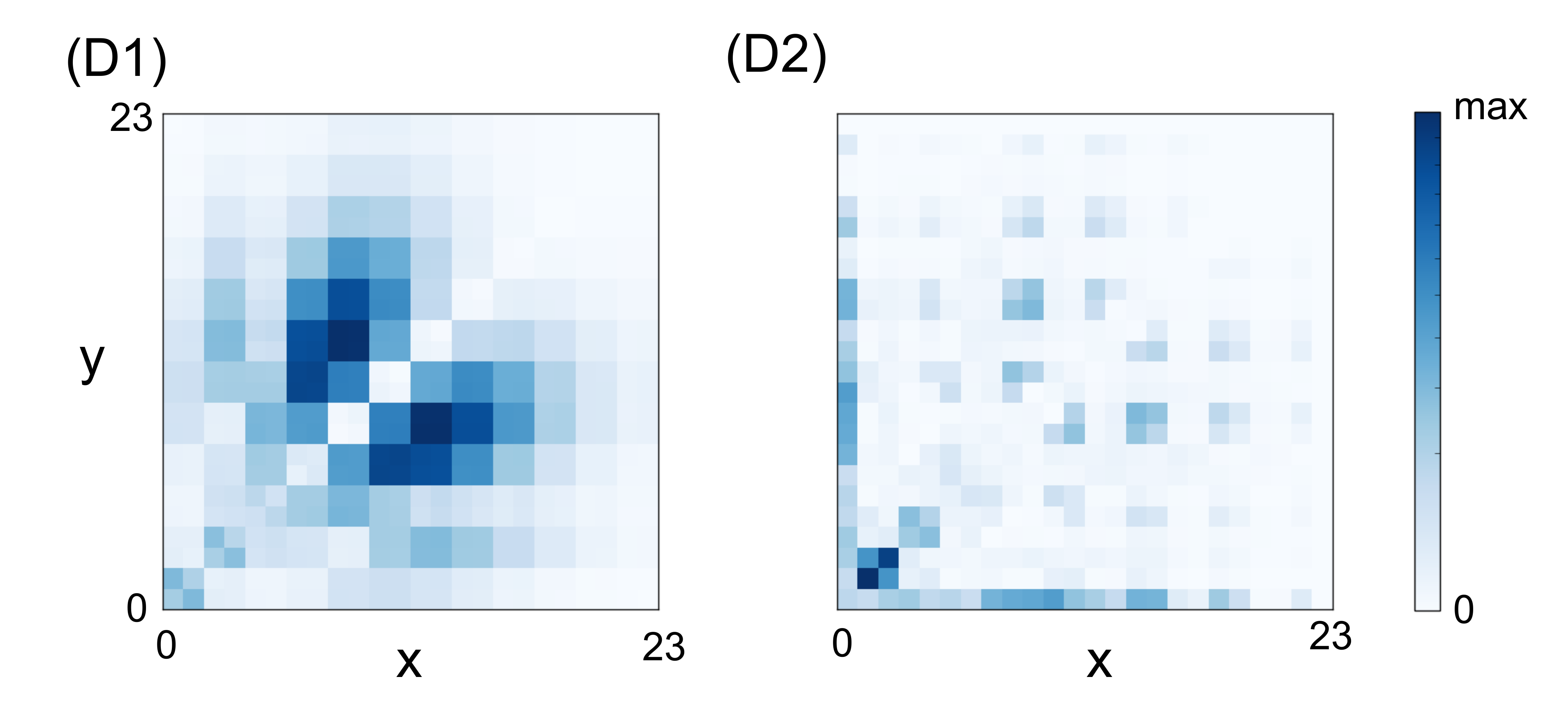}
\caption{Modulus of the two-body wavefunction after time
  evolution of the initial states in Eqs.~(\ref{but1}, \ref{but2}) for (D1) 
  dimerization $D1$ and (D2) dimerization $D2$ at time $t=75 \hbar/J_1$. In
  these simulations $U=1.5J_1$, $J_2=0.1J_1$ and $L=24$ sites.}
\label{fig:tEvol-but}
\end{figure}

For that reason, the time evolution, shown in
Fig.~\ref{fig:tEvol-but}, presents two drastically different
behaviours in the two dimerizations: in $D1$ a two-body scattering
pattern develops, which covers the central part of the lattice leaving
the density on the diagonal suppressed due to interactions; in $D2$
the two-body wavefunction presents an admixture of two free scattering
particles and type II edge-scattering states.

\vspace*{0.5cm}

%%%% CONCLUSIONS %%%%%%

\section{Conclusions}

In conclusion, we have studied theoretically the
rich two-particle physics stemming from the interplay of local interactions with  
non-trivial single-particle topology. To this aim, we have considered two particles in the paradigmatic 
one-dimensional Su-Schrieffer-Heeger dimerized lattice.
We have proposed an experimentally realistic system, based on state-of-the-art coupled optical fiber technology, where the
two-body physics in the SSH model can be quantum simulated in real time and real space. 
Beyond being able of revealing the different scattering,
bound and edge bound states in finite geometries, experiments along
the suggested lines have the potential of becoming a textbook
illustration of the Fano-Feshbach resonance scattering effect.

One of our major conceptual results resides 
in the evidence that interactions, in spite of being local, can affect the boundary 
conditions over more than one single lattice site. Such kind of effects are expected to be even more relevant 
in the presence of non-local interactions. Hence, the most straightforward extension
of the present work regards 
the inclusion of nearest-neighbor interactions \cite{EPJST}. 

Our work provides a first important progress in the understanding of two-body physics
in systems with topological properties. In the future, it would be interesting
to investigate models in higher dimensions and different geometries, where
symmetries other than the chiral one are relevant for the existence of topological states.

%%%%% ACKNOWLEDGEMENTS %%%%%

\section{Acknowledgements}

The authors thank M.~Burrello, P.~\"Ohberg,
C.~Ortix, T.~Ozawa and H.~Price for interesting discussions.
A.R. acknowledges support from the Alexander von Humboldt foundation
and W.~Zwerger for the kind hospitality at the TUM. This work was
supported by ERC through the QGBE grant, by the EU-FET Proactive grant
AQuS, Project No.  640800 and by Provincia Autonoma di Trento.  

\emph{Note added}. In the final stage of preparation of this work, we
became aware of a similar and complementary investigation of the two
particle SSH model by Gorlach and Poddubny \cite{Gorlach2016}.

%%%%%% APPENDICES %%%%%%%%

\appendix

%%%%% EFFECTIVE THEORIES %%%%%%

%%%%%%%%%%%%%%%%%%%%%%%%%%%%%%%%
\section{Effective theories}
%%%%%%%%%%%%%%%%%%%%%%%%%%%%%%%%
\label{eff}

It this work, we have made extensive use of effective models to
describe two bosonic particles in a dimerized lattice governed by the
Hamiltonian $H =H_{J_1} + H_{J_2} + H_U$, where $H_{J_1}$ and
$H_{J_2}$ are the strong- and weak-tunneling Hamiltonians, and $H_U$
accounts for onsite interactions. In this section, we provide the
details of their derivation.

Consider a Hamiltonian $H = H_0 + V$. Let us label the set of
eigenstates of $H_0$ as $\{\alpha\}=\{|\alpha,m\ket\}$. Here, the
index $\alpha$ indicates a manifold of states (for instance, states
close in energy to each other that are gapped from the rest of the
other states), and $m$ labels the states inside the manifold. Be $V$ a
perturbation that weakly couples the manifold $\{\alpha\}$ to the
manifold $\{\beta\}$ of the remaining eigenstates of $H_0$. Including
$V$ at second order perturbation theory, as shown in \cite{Cohen}, one
can obtain an effective Hamiltonian $H^{\alpha}_\textrm{eff}$ that
describes manifold~$\{\alpha\}$
\begin{align}
\mv{\alpha, m|H^{\alpha}_\textrm{eff}|\alpha, n} = E_{\alpha,m}
\delta_{m,n} + \mv{\alpha,m | V | \alpha,n} +\\ +\f 1
2\sum_{k,\beta\neq\alpha} \mv{\alpha,m | V | \beta,k} \mv{\beta,k | V
  | \alpha,n} \times \nn\\ \times \left[
  \f{1}{E_{\alpha,m}-E_{\beta,k}} + \f{1}{E_{\alpha,n}-E_{\beta,k}}
  \right]\,,\nn
\end{align}
where $E_{\alpha,m}$ are the eigenvalues of $H_0$ relative to the eigenstate $|\alpha,m\ket$.

%%%%%%%%%%%%%%%%%%%%%%%%%%%%%%%%
\subsection{Strong dimerization}
\label{app_strong_dim}
%%%%%%%%%%%%%%%%%%%%%%%%%%%%%%%%

\begin{figure}[!tbp]
  \includegraphics[width=0.75\columnwidth]{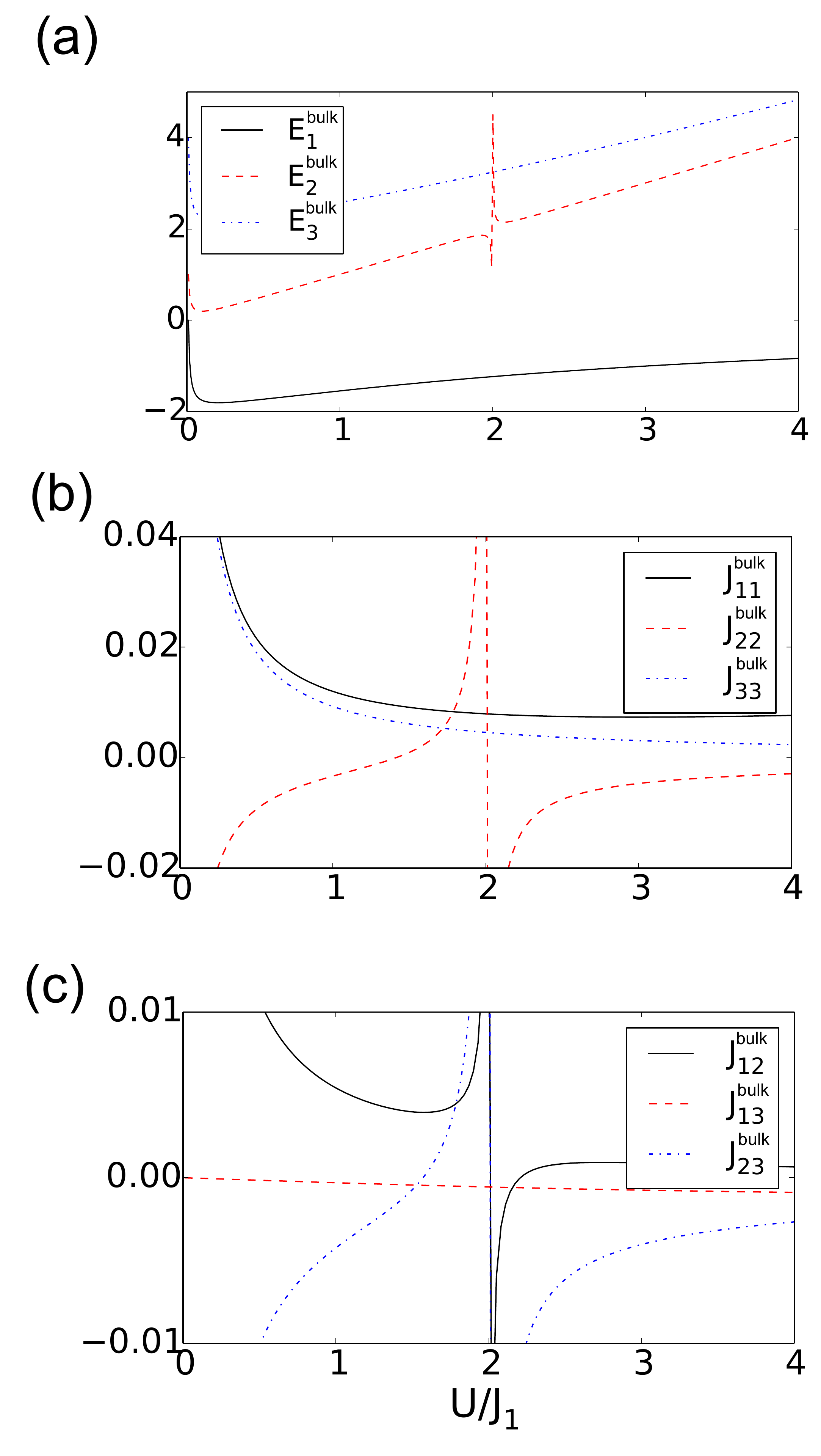}
  \caption{Bulk parameters $E^{\textrm{bulk}}_\alpha$ and $J^{\textrm{bulk}}_{\alpha\beta}$ of the
    effective model in Eq.~(\ref{effHam2_SM}) for $J_2=0.1J_1$, as a
    function of $U$. The detailed legend can be found in the figure. }
\label{fig:par}
\end{figure}

In the limit of strong dimerization $J_2 \ll J_1, U$, we identify $H_0
= H_{J_1}+H_U=\sum_i H^{\textrm{cell}}_i$. Different lattice cells are
decoupled and each cell is described by the strong-link Hamiltonian
$H^{\textrm{cell}}_i$, which in the two-particles basis $|A_i,A_i\ket$,
$|A_i,B_i\ket$ and $|B_i,B_i\ket$ takes the form
\be
H^{\textrm{cell}}_i
=
\begin{pmatrix}
U & -\sqrt{2} J_1 & 0 \\
-\sqrt{2} J_1 & 0 & -\sqrt{2} J_1\\
0 & -\sqrt{2} J_1 & U
\end{pmatrix}\,.
\label{H_cell}
\ee
Its eigenvectors, provided in Eqs.~(\ref{d1}-\ref{d3}),
have respectively energy 
\begin{eqnarray}
  \epsilon_{1} &=& \frac{1}{2}\left( U - \sqrt{16J_1^2 + U^2} \right),
  \label{e1} \\ 
\epsilon_2 &=& U, \label{e2} \\  
\epsilon_{3} &=& \frac{1}{2} \left( U + \sqrt{16J_1^2 + U^2} \right). \label{e3} 
\end{eqnarray}
The states in manifold
$\{\alpha\}=\{|d_{1,i}\ket,|d_{2,i}\ket,|d_{3,i}\ket\}$ are coupled
through $H_{J_2}$ in a non-trivial manner via the manifold of virtual
states $\{\beta\}$ - also eigenstates of $H_0$. For PBC, manifold
$\{\beta\}$ is formed by states of one particle in a cell $i$ and one
particle in a cell $j$, with $i\neq j$. There are four possible sets
of states
\begin{align}
|\psi^I_{ij}\ket &= \f{1}{\sqrt{2}} \left( |A_i\ket + |B_i\ket \right) \otimes \f{1}{\sqrt{2}} \left( |A_j\ket + |B_j\ket \right)\,,\\
|\psi^{II}_{ij}\ket &= \f{1}{\sqrt{2}} \left( |A_i\ket - |B_i\ket \right) \otimes \f{1}{\sqrt{2}} \left( |A_j\ket - |B_j\ket \right)\,,\\
|\psi^{III}_{ij}\ket &= \f{1}{\sqrt{2}} \left( |A_i\ket + |B_i\ket \right) \otimes \f{1}{\sqrt{2}} \left( |A_j\ket - |B_j\ket \right)\,,\\
|\psi^{IV}_{ij}\ket &= \f{1}{\sqrt{2}} \left( |A_i\ket - |B_i\ket \right) \otimes \f{1}{\sqrt{2}} \left( |A_j\ket + |B_j\ket \right)\,,
\end{align}
with energies, respectively, $E^{I} = -2J_1$, $E^{II}=2J_1$, $E^{III}=E^{IV}=0$.
Up to second order in perturbation $V=H_{J_2}$, one finds the
effective Hamiltonian (see Eq.~(\ref{effHam2}))
\be
H_{\textrm{eff}} = \sum_{i,\alpha} E_{\alpha,i}\, d^\dag_{\alpha,i} d_{\alpha,i} + \sum_{\mv{i,j}}\sum_{\alpha,\beta} J_{\alpha\beta}^{ij}\, d^\dag_{\alpha,i} d_{\beta,j}\,,
\label{effHam2_SM}
\ee
containing renormalized onsite dimer energies and intra- and
inter-dimer nearest-neighbor hopping.  In general, coefficients
$E_{\alpha,i}$ and $J_{\alpha\beta}^{ij}$ have a quite involved
analytical form. The values of the parameters for $J_2=0.1 J_1$
as a function of $U$ are shown in Fig.~\ref{fig:par} when
$\mv{i,j}$ are in the bulk of the lattice where
no edge effects are involved. The divergencies
at $U=2J_1$ are the indication of the crossing of $d_2$ with the
higher type I continuum.

In most regimes, the different bound states are far away in
energy from each other and the coupling among them turns out to be weak.
Even if better quantitative predictions for the bound states bands 
can be obtained by including all the terms, a decoupling of the different
bound states, namely considering for each bound state $d_\alpha$ only
the parameters $J_{\alpha\alpha}^{ij}$ and $E_{\alpha,i}$, still provides
an excellent agreement. In that
case, explicit analytical forms can be provided at least for the
simpler case of state $d_2$:
\begin{align}
J_{22}^{\textrm{bulk}} &= -\f{J_2^2}{U} \f{2J_1^2-U^2}{4J_1^2-U^2}\,, \\
E_2^{\textrm{bulk}} &= U - 2 J_{22}^{\textrm{bulk}} \,. 
\end{align}
For OBC, bulk and edge parameters differ: in $D1$, due to the missing
coupling either on the right-hand or left-hand side, one gets a
different renormalization of the onsite energy at the first and last
cells:
\be
E_{2}^{\textrm{edge,D1}} = U - J_{22}^{\textrm{bulk}}\,, \ee
while the effective hopping parameter is equal at the edges as in the bulk.

Different parameters describe dimerization $D2$, due to the presence
of half cells (single lattice sites) at the chain edges.
The effective tunneling coupling between the first (last) lattice
site to the first (last) complete lattice cell and the energy offset
of the first (last) lattice sites are
\begin{eqnarray}
J_{22}^{\textrm{edge,D2}} &=& \f{J_2^2}{U} \f{\sqrt 2 U^2}{J_1^2-U^2}\,,\\
E_{2}^{\textrm{edge,D2}} &=& U - \sqrt 2 J_{22}^{\textrm{edge,D2}}\,.
\end{eqnarray}
Finally, the first ($i=1$) and last ($i=L-1$) complete cells feel a
resulting energy shift given by

\be E_{2,1} = E_{2,L-1} =U+
\f{J_2^2}{U}\f{2J_1^4-7J_1^2U^2+2U^4}{4J_1^4-5J_1^2U^2+U^4}\, .
\ee

%%%%%%%%%%%%%%%%%%%%%%%%%%%%%%%%
\subsection{Strong-interaction limit}
\label{app_strong_int1}
%%%%%%%%%%%%%%%%%%%%%%%%%%%%%

In the strong-interaction limit $U\gg J_1,J_2$, the two
  $d_{2,3}$ narrow bound-state bands are well separated from the rest
of the spectrum, as one can deduce from Eqs.~(\ref{e2},\ref{e3}) and
Figs.~\ref{spectrum_OBC_U}(D1,D2).  We choose linear combinations of 
 these higher repulsive
bound states to constitute the manifold $\{\alpha\}$ for which we
write the effective theory. This corresponds to consider $H_0 = H_U$
and the subspace of onsite doublons $\{\alpha\} = \{|A_i,A_i \ket ,
|B_i,B_i \ket \}\equiv \{ d_{A,i}^\dag|0\ket, d_{B,i}^\dag|0\ket\}$,
with energy $E_\alpha = U$.

The virtual states at energy $E_\beta = 0$ form manifold $\{\beta\} =
\{ |A_i,A_j \ket , |B_i,B_j \ket \, |A_m,B_n \ket \}$ with $i\neq j$
and $\forall m,n$.  The coupling is provided by $V=H_{J_1} +
H_{J_2}$. Up to second order in $J_{\sigma}$, only $|A_i,B_{i}\ket$
and $|A_i,B_{i-1}\ket$ contribute, corresponding to nearest-neighbor
virtual hopping processes.

One obtains an effective single-particle Hamiltonian for the on-site
doublons that reads
\begin{align}
\label{effHam1}
H_\textrm{eff} =& \f{2J_1^2}{U} \sum_i d^\dag_{A,i}d^{}_{B,i} + \f{2J_2^2}{U} \sum_i d^\dag_{A,i+1}d^{}_{B,i}+\textrm{H.c.} \nn \\
& +  \left[ U + \f{2}{U} \left( J_1^2 + J_2^2 \right) \right] \sum_i \left( d^\dag_{A,i}d^{}_{A,i} + d^\dag_{B,i}d^{}_{B,i} \right)\,.
\end{align}
The first line clearly shows that the effective model of the bound
states is a SSH model with renormalized hopping coefficients
\be
J_{1,2}^\textrm{eff} = -\frac{2J_{1,2}^2}{U}\, 
\ee
and effective on-site energies
\be \epsilon_{\textrm{bulk}} = U +
\f{2}{U} \left( J_1^2 + J_2^2 \right)\,,
\ee
which contains the binding energy $U$ and an on-site energy shift. The
latter is generated by similar second-order processes as the ones
occurring for the hopping terms: a virtual breaking of the doublon to
the left and to the right. However, in the presence of OBC, at the
edges only one of these two processes will be present, leading to a
different on-site energy at the edge with respect to the bulk
\be
\epsilon_{\textrm{edge}} = \epsilon_{\textrm{bulk}} +\Delta E_\sigma,
\ee
with $\Delta E_\sigma =- 2J_{3-\sigma}^2/U$, depending on the
dimerization $D\sigma$ (with $\sigma=1,2$).

\begin{figure}[!tbp]
  \includegraphics[width=0.8\columnwidth]{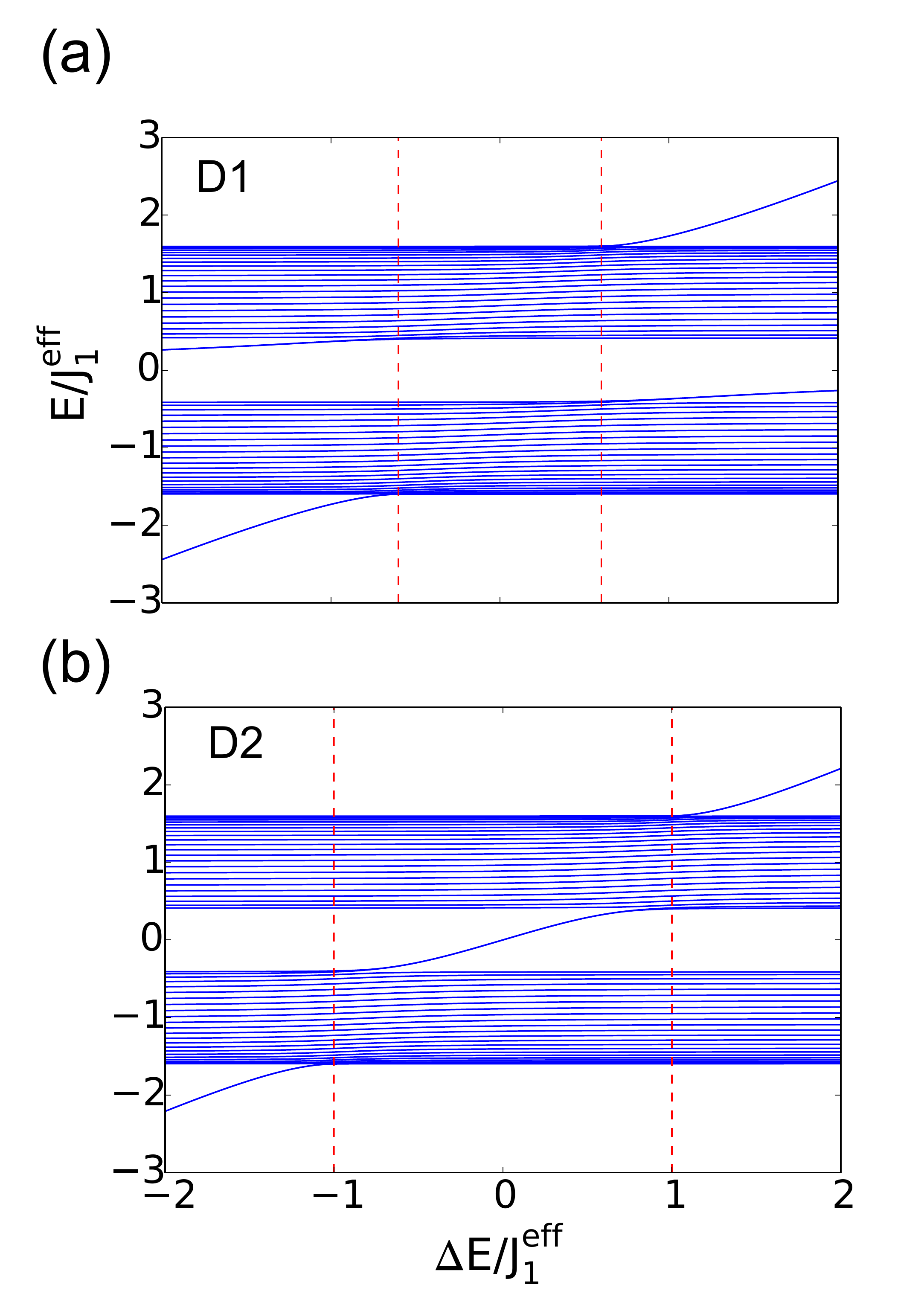}
\caption{Single-particle SSH spectrum of a finite chain with $48$
  sites for $J^{\textrm{eff}}_2=0.6 J^{\textrm{eff}}_1$ as a function of an arbitrary on-site energy shift $\Delta
  E$ for (a) dimerization $D1$ and (b) dimerization $D2$.  As discussed in
  the text, this single-particle model
  effectively describes the bound-states physics in the
  strong-interaction limit. The red lines are the critical values of
  $\Delta E$ for the existence of Tamm or in-gap edge states (see
  text).  }
  \label{tamm}
\end{figure}

This effective Hamiltonian provides a generalization of the Tamm physics
to the SSH model, which is summarized in Fig.~\ref{tamm}.  
For varying $\Delta E$, the energy of the Tamm-like states lies
above/below the bands and it depends linearly on $\Delta E$ when 
$|\Delta E|$ is sufficiently large. They appear for $|\Delta E| > J_2^\textrm{eff}$ in
dimerization $D1$ and for $|\Delta E| > J_1^\textrm{eff}$ in
dimerization $D2$.  In-gap states between the 
bands appear for both dimerizations, 
but they exist in $D1$ when $|\Delta E| > J_2^\textrm{eff}$ and
in $D2$ when $|\Delta E|<J_1^\textrm{eff}$. 

The in-gap edge states obtained when considering $\Delta E$ as 
tunable parameter, have a topological origin. 
The peculiar feature of the SSH model is indeed the presence of zero-energy edge
states in dimerization $D2$ that are topologically protected by chiral
symmetry. However, chiral symmetry is broken by the presence of the
off-set $\Delta E$, as proven below. As a consequence, topological edge 
states in $D2$ are not protected anymore and, for moderate $\Delta E$, shift away from zero energy, 
as discussed in the paragraph above and shown in Fig.~\ref{tamm}(b). 
Also the in-gap states in dimerization $D1$ have a similar topological origin. Indeed,
one can observe from Fig.~\ref{tamm}(a) that these states asympotically tend to
zero energy for infinitely large values of $\Delta E$. In this limit, the 
system behaves as an ideal $D2$ dimerized lattice with $L-2$ sites, which must possess
zero-energy edge states.

In the specific case of the effective model derived in this section, 
we have found that $|\Delta E_1| =
J_2^\textrm{eff}$ and $|\Delta E_2 |= J_1^\textrm{eff}$, which implies
that we are exactly at the critical values of $\Delta E$ (red lines in
Fig.~\ref{tamm}) for which neither Tamm nor in-gap states can exist.

We now prove the breaking of chiral symmetry in the effective model.
Let us consider a finite chain with $2L$ sites in dimerization
$D2$. This corresponds to $L-1$ full cells and two single lattice
sites at the boundaries.
After defining the vector $\b d =
(d_{A,1},\dots,d_{A,L},d_{B,0},\dots,d_{B,L-1}$), the Hamiltonian
takes the form (up to an overall irrelevant energy shift that we drop
in the discussion below)
\be H_{\textrm{eff}}= \b d^\dagger \mathcal
H_{\textrm{eff}} \b d = \b d^\dagger \left(
\begin{array}{cc}
\mathcal H_{11} & \mathcal H_{12} \\
\mathcal H_{12}^\dagger & \mathcal H_{22}
\end{array}
\right) \b d \,.
\ee

While $\mathcal H_{12}$ is a $L\times L$ matrix that contains the
coupling between neighboring sites and has the form
\be
\mathcal H_{12}=
\left( 
\begin{array}{ccccc}
 -J_1^\textrm{eff} &  &  &  & 0\\ [6pt]
-J_2^\textrm{eff} & -J_1^\textrm{eff} & & & \\
 & \ddots & \ddots &  & \\
 &  & \ddots & -J_1^\textrm{eff} &  \\ [6pt]
 0 &  &  & -J_2^\textrm{eff} & -J_1^\textrm{eff}
\end{array}
\right)\,,
\ee
$\mathcal H_{11}$ and $\mathcal H_{22}$ are $L\times L$ diagonal
matrices, namely $\mathcal H_{11} = \textrm{diag}(0 ,\cdots, 0,\Delta
E_2)$ and $\mathcal H_{22} = \textrm{diag}(\Delta E_2,0,\cdots, 0)$ that
describe the on-site energy shift, respectively, of the left and right
edge of the chain.
If $\Delta E_2$ were zero, the diagonal blocks $\mathcal H_{11}$ and
$\mathcal H_{22}$ would vanish. Hence, the Hamiltonian could be written in the
chiral-symmetric form $\mathcal H_\textrm{eff}^{(0)} = \begin{pmatrix}
  \b 0 & \mathcal H_{12}\\ \mathcal H_{12}^\dagger & \b
  0 \end{pmatrix}$. In fact, in this case, the operator $\mathcal C =
\sigma_z \otimes \mathcal{I}$ provides a chiral symmetry such that
$\{\mathcal H_\textrm{eff}^{(0)}, \mathcal C\} = 0$. However, since in our case $\Delta
E_2 \neq 0$, diagonal blocks appear in $\mathcal
H_{\textrm{eff}}$. Chiral symmetry is therefore broken and zero-energy edge
states are not protected \cite{Ryu2002}.

\vspace*{0.2cm}

%%%%%%%%%%%%%%%%%%%%%%%%%%%%%%%%
\subsection{Effective theory for the $d_{NN}$ state}
\label{app:dnn}
%%%%%%%%%%%%%%%%%%%%%%%%%%%%%%%%

We develop an effective theory in the strong-dimerization limit $J_2
\ll J_1$ to qualitative explain the existence of the $d_{NN}$ bound
state. The basis for the effective theory is given by the subspace of states $|S_{ij}\ket \sim (|A_i\ket -
|B_i\ket ) \otimes ( |A_{j}\ket - |B_{j}\ket )$ with $i$, $j$
arbitrary cell indices. For $i\neq j$, the states $|S_{ij}\ket$ span the upper type I scattering continuum. For
$i=j$, $|S_{ii}\ket = (|d_3(U=0)\ket$, which can be considered a fairly
good approximation for $d_3$ up to $U\le 2J_1$.  At first order in
$J_2$, the energies of states $|S_{ij}\ket$ are $E_{ij}=2J_1+\delta_{ij}
U/2$. The single-particle hopping amplitude in this subspace is given
by $J_2/2$. When $U$ is approaching $2 J_1$ (but sufficiently far to
be off-resonant with the type I scattering states), state $d_2$ become
closer in energy to the $S_{ij}$ manifold. Then, the energy of the
states $|S_{ij}\ket$ with $i=j\pm 1$ is not simply given by $2 J_1$
but it is renormalized by second-order processes mediated by the
virtual state $d_2$.
The energy shift is given by

\be \Delta E_{NN} = 2\times \f{J_2^2}{4} \f{1}{2J_1-U} \ee
and provides an effective nearest-neighbor attractive (repulsive)
interaction when $U>2J_1$ ($U<2J_1$). Therefore, for $U>2J_1$
($U<2J_1$) we expect a bound state above (below) the continuum. The
nearest-neighbor interaction is very weak compared to the bandwidth
$2J_2$ of the scattering states. This explains the
appearance of the $d_{NN}$ state for a limited set of momenta close to
$K=\pi$ \cite{Valiente2009}. Moreover, for increasing values of $U$
the attraction becomes weaker and weaker, leading to a progressive
disappearance of the $d_{NN}$ state. These properties have all been
observed numerically (see Fig.~\ref{fig:PBC}(a-b)).

%%%%%%%%%%%%%%%%%%%%%%%%%%%%%%%%
\subsection{Reduced theory for the avoided crossing}
\label{app:reduced}
%%%%%%%%%%%%%%%%%%%%%%%%%%%%%%%%

Using the formalism presented in \ref{app_strong_dim}, we were able to
describe the narrow bound states bands. However, the hybridization
between bound states and type I scattering states presented in
Fig.~\ref{fig:crossing} has to be accounted for with a different
model, including type II scattering states as real rather than virtual states.

The Hilbert space of the reduced theory for the avoided crossing is
provided by the set of states $\{|\psi^{\textrm{red}}\ket \}= \{
|d_{2,0}\ket\,,|\psi^{l}_{i}\ket\,, |\psi^{\textrm{loc}}\ket\,,
|\psi^{r}_{i}\ket\,, |d_{2,L}\ket \}$:
\begin{align}
|d_{2,0}\ket &= -|B_0, B_0\ket\,, \nn\\
|\psi^{l}_{i}\ket &= |B_0\ket \otimes \left(|A_i\ket - |B_i\ket \right)/\sqrt 2\quad i=1,\cdots,L-1\,, \nn \\
|\psi^{\textrm{loc}}\ket &= |B_0 A_L\ket\,, \nn \\
|\psi^{r}_{i}\ket &= |A_L\ket \otimes \left(|A_i\ket - |B_i\ket \right)/\sqrt 2 \quad i=1,\cdots,L-1 \,, \nn \\
|d_{2,L}\ket &= |A_L, A_L\ket \nn\,.
\end{align}
One therefore constructs the reduced Hamiltonian 
\be
\mv{\psi^{\textrm{red}}_\alpha |\mathcal H^{\textrm{red}}|\psi^{\textrm{red}}_\beta} \equiv \mv{\psi^{\textrm{red}}_\alpha |H|\psi^{\textrm{red}}_\beta},
\ee
that reads
\begin{widetext}    

\be
\mathcal H^{\textrm{red}}=
\left( 
\begin{array}{c|ccccc|c|ccccc|c}
 U & J_2 & & & & & & & & & & & \\ \hline 
 J_2& J_1 & -\f{J_2}{2} &  &  & & & & & & & &\\
 &-\f{J_2}{2} & J_1 & \ddots & & & & & & & & &\\ [0pt]
 & & \ddots & \ddots & \ddots & & & & & & & &\\ [0pt]
 & &  & \ddots & J_1 & -\f{J_2}{2} & & & & & & &\\ [0pt]
 &  &  &  & -\f{J_2}{2} & J_1 & \f{J_2}{\sqrt 2} & & & & & &\\ \hline
 & & & & & \f{J_2}{\sqrt 2} & 0 & -\f{J_2}{\sqrt 2} & & & & &\\ \hline
 & & & & & & -\f{J_2}{\sqrt 2} & J_1 & -\f{J_2}{2} & & & &\\
 & & & & & & & -\f{J_2}{2} & J_1 & \ddots & & &\\
 & & & & & & & & \ddots & \ddots & \ddots & &\\
 & & & & & & & & & \ddots & J_1 & -\f{J_2}{2} & \\
 & & & & & & & & & & -\f{J_2}{2} & J_1 & J_2 \\ \hline
 & & & & & & & & & & & J_2 & U
\end{array}
\right)\,.
\ee
\end{widetext}
The diagonalization of $H^{\textrm{red}}$ shows a very good agreement
with the complete spectrum obtained by exact diagonalization, as shown
in Fig.~\ref{fig:crossing}.
\bibliography{BiblioSSH}

\end{document}